\documentclass[reprint,prl,aps,superscriptaddress,twocolumn]{revtex4}
\usepackage{times}
\usepackage{graphicx}
\usepackage{amsmath, braket, amsfonts}
\usepackage{amssymb}
\usepackage{physics}
\usepackage{sidecap}
\usepackage{bm, color, ulem}
\usepackage{xcolor}
\usepackage{ragged2e}
\usepackage{wrapfig,lipsum,booktabs}
\usepackage{multirow}
\usepackage{tabularx}

\usepackage[separate-uncertainty=true]{siunitx}

\newcommand{\be}{\begin{equation}}
\newcommand{\ee}{\end{equation}}

\DeclareUnicodeCharacter{03BD}{{$\nu$}}

\newcommand{\vect}[1]{{\mathbf{#1}}}
\newcommand{\bq}{\vect{q}}

\newcommand{\bR}{\vect{R}}

\newcommand{\RPA}

\makeatletter
\def\maketitle{
\@author@finish
\title@column\titleblock@produce
\suppressfloats[t]}
\makeatother

\begin{document}
\title{Energy gap of the even-denominator fractional quantum Hall state in bilayer graphene}
\author{Alexandre Assouline}
\affiliation{Department of Physics, University of California at Santa Barbara, Santa Barbara CA 93106, USA}
\author{Taige Wang}   
\affiliation{Department of Physics, University of California, Berkeley, California 94720, USA}
\affiliation{Material Science Division, Lawrence Berkeley National Laboratory, Berkeley, California 94720, USA}
\author{Haoxin Zhou}   
\affiliation{Department of Physics, University of California at Santa Barbara, Santa Barbara CA 93106, USA}
\author{Liam A. Cohen}   
\affiliation{Department of Physics, University of California at Santa Barbara, Santa Barbara CA 93106, USA}
\author{Fangyuan Yang}   
\affiliation{Department of Physics, University of California at Santa Barbara, Santa Barbara CA 93106, USA}
\author{Ruining Zhang}   
\affiliation{Department of Physics, University of California at Santa Barbara, Santa Barbara CA 93106, USA}

 \author{Takashi Taniguchi}
 \affiliation{International Center for Materials Nanoarchitectonics,
 National Institute for Materials Science,  1-1 Namiki, Tsukuba 305-0044, Japan}
 \author{Kenji Watanabe}
 \affiliation{Research Center for Functional Materials,
 National Institute for Materials Science, 1-1 Namiki, Tsukuba 305-0044, Japan}
 \author{Roger S. K. Mong}
\affiliation{Department of Physics and Astronomy,
University of Pittsburgh, Pittsburgh, PA 15260, USA}
 \author{Michael P. Zaletel}   
\affiliation{Department of Physics, University of California, Berkeley, California 94720, USA}
\affiliation{Material Science Division, Lawrence Berkeley National Laboratory, Berkeley, California 94720, USA}

\author{Andrea F. Young}
\email{andrea@physics.ucsb.edu}
 \affiliation{Department of Physics, University of California at Santa Barbara, Santa Barbara CA 93106, USA}
\date{\today}

\begin{abstract}
Bernal bilayer graphene hosts even denominator fractional quantum Hall states thought to be described by a Pfaffian wave function with nonabelian quasiparticle excitations.   
Here we report the quantitative determination of fractional quantum Hall energy gaps in bilayer graphene using both thermally activated transport and by direct measurement of the chemical potential. 
We find a transport activation gap of $\SI{5.1}{K}$ at $B = \SI{12}{T}$ for a half-filled $N=1$ Landau level, consistent with density matrix renormalization group calculations for the Pfaffian state. 
However, the measured thermodynamic gap of $\SI{11.6}{K}$ is smaller than theoretical expectations for the clean limit by approximately a factor of two.  
We analyze the chemical potential data near fractional filling within a simplified model of a Wigner crystal of fractional quasiparticles with long-wavelength disorder, explaining this discrepancy.  
Our results quantitatively establish bilayer graphene as a robust platform for probing the non-Abelian anyons expected to arise as the elementary excitations of the even-denominator state.
\end{abstract}

\maketitle

Non-Abelian anyons\cite{moore_nonabelions_1991} are thought to enable fault tolerant topological quantum bits through their non-trivial braiding statistics\cite{nayak_non-abelian_2008}. 
In an ideal scenario, the error rate of such qubits is limited only by the  density of thermally excited quasiparticles present in the system. Such processes---analogous to quasiparticle poisoning in superconducting qubits---are exponentially suppressed at low temperature by an Arrhenius law, $n_{\mathrm{qp}} \propto \exp\left(-\Delta_{\mathrm{qp}}/2 k_B T\right)$, where $\Delta_{\mathrm{qp}}$ is the energy gap for non-Abelian quasiparticles and $ T$ is temperature. The energy gap is thus a key figure of merit for candidate nonabelian states. 
According to numerical calculations\cite{morf_transition_1998,rezayi_landau-level-mixing_2017}, nonabelian ground states are the leading candidates to describe the even denominator fractional quantum Hall (FQH) states observed in the second orbital Landau level of single-component systems such as GaAs quantum wells\cite{willett_observation_1987}. 
While these numerical results are thought to be reliable, the small energy gaps measured for these states in GaAs\cite{kumar_nonconventional_2010,watson_impact_2015,chung_ultra-high-quality_2021} have hampered experimental efforts to directly probe nonabelian statistics via fusion and braiding of individual quasiparticles.  

Within the simplest model of bilayer graphene, the $N=0$ and $N=1$ orbital levels are both pinned to zero energy\cite{mccann_landau-level_2006}. Combined with the spin- and valley degeneracies native to graphene quantum Hall systems\cite{dean_fractional_2020}, this produces an eight-fold degeneracy---a seemingly inauspicious arena for the single-component physics of nonabelian FQH states. 
However, as a wealth of experimental work has shown, all of these degeneracies are lifted by the combination of electronic interactions and the applied displacement field\cite{feldman_broken-symmetry_2009,martin_local_2010,dean_boron_2010,lee_chemical_2014,kou_electron-hole_2014,ki_observation_2014,maher_tunable_2014,hunt_direct_2017,huang_valley_2022,zibrov_tunable_2017,li_even_2017}.
In particular, broad domains of density and displacement field are characterized by partial filling of a singly degenerate $N=0$ or $N=1$ Landau level. 
In the $N=1$ regime, an incompressible state is observed at half-integer filling\cite{ki_observation_2014,zibrov_tunable_2017,li_even_2017,huang_valley_2022}, which calculations show should be described by a nonabelian Pfaffian ground state\cite{apalkov_stable_2011, papic_topological_2014, balram_transitions_2022,zibrov_tunable_2017}. Prior measurements of energy gaps have found activation gaps as large as \SI{1.8}{K} at $ B = \SI{14}{T}$; however, precise comparisons of activation and thermodynamic gaps to theoretical expectations have not been previously reported.

\begin{figure*}[ht]
    \centering
    \includegraphics[width = 180mm]{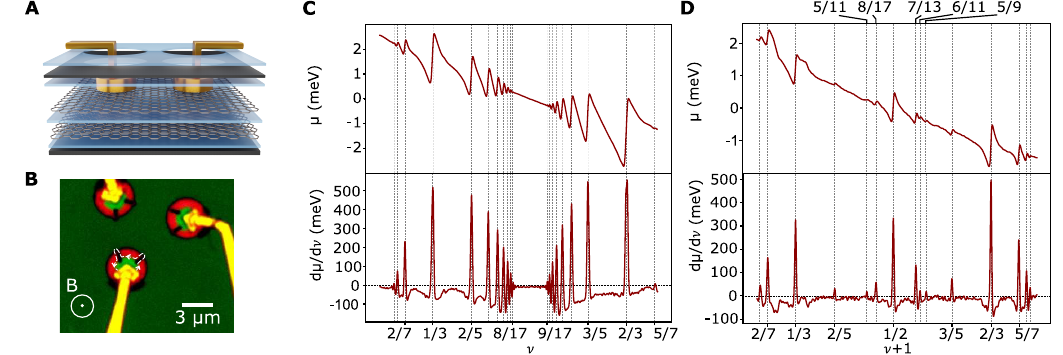}
    \caption{\textbf{Chemical potential and inverse compressibility of bilayer graphene fractional quantum Hall states.}
    \textbf{(A)} Device schematic showing the hBN layers (blue), top and bottom graphite gates (dark grey), monolayer graphene detector layer connected to Corbino contacts, and bilayer graphene sample layer. 
    \textbf{(B)} Optical image of the Corbino contacts to the monolayer graphene detector.
    White dashed lines show the trajectory of a chiral edge state along trenches etched through the device, which ensures contact between the metal and dual gated sample bulk.  
    \textbf{(C)} The top panel shows the measured $\mu$ at $B=\SI{13.8}{T}$ and $T=\SI{50}{mK}$ in the partially filled N=0 level spanning $0<\nu<1$.  The bottom panel shows the inverse compressibility, $d\mu/d\nu$, calculated by numerically differentiating the data in the top panel. 
    \textbf{(D)} The same as C, but for the partially  filled N=1 orbital Landau level spanning $-1<\nu<0$. 
    }
    \label{fig:1}
\end{figure*}

Here we report energy gaps for both odd- and even-denominator FQH states in bilayer graphene using both transport and chemical potential measurements. 
Thermally activated transport measures the energy cost of creating a physically separated quasiparticle-quasihole pair.  
We measure activated transport using a Corbino-like geometry\cite{polshyn_quantitative_2018,zeng_high-quality_2019}, which directly  probes the conductivity of the gapped, insulating bulk.  
Chemical potential measurements record a jump at incompressible filling factors known as the thermodynamic gap, which---in the clean limit---measures the difference between adding charge $\pm e$ to the gapped system. We measure the thermodynamic gap using a direct-current charge sensing technique based on a double-layer device\cite{eisenstein_compressibility_1994,yang_experimental_2021}.  
Combining these techniques, we find several new features, including weak FQH states at $\nu=5/11$, $\nu=6/11$ and $\nu=5/9$ of a partially filled N=1 Landau level. 
Moreover, both schemes show an energy gap for a half-filled single component Landau level that is several times larger than reported to date for a candidate nonabelian state in any system
\cite{gammel_ultralow-temperature_1988,kumar_nonconventional_2010,watson_impact_2015,falson_even-denominator_2015,zibrov_tunable_2017,li_even_2017,shi_odd-_2020,chung_ultra-high-quality_2021}.  
Notably, these measurement schemes effectively average over $\sim\SI{10}{\mu m^2}$ sized areas, a testament to the exceptional uniformity of the electron gas in bilayer graphene.  

Fig. 1A shows a schematic of the experimental geometry used to measure the chemical potential $\mu$. A graphene bilayer hosting the FQH system of interest is  separated by a \SI{62}{nm}-thick hexagonal boron nitride (hBN) dielectric from a graphene monolayer that functions as a sensor.  
Both layers are encapsulated by additional hBN dielectrics and graphite gates, creating a four plate geometry that allows independent control of the carrier density on both the monolayer detector and bilayer sample layer. 
We measure Corbino transport in the detector layer, where a FQH state functions as a sensitive detector of the local potential.  An optical image of the Corbino contacts is shown in Fig. 1B. 
As described in detail in the supplementary information, monitoring transport in the sensor layer allows us to precisely determine $\mu$ of the bilayer sample. An advantage of our technique is that it avoids finite-frequency modulation of the carrier density, allowing us to accommodate charge equilibration times as large as a second. 

Figs. \ref{fig:1}C-D show $\mu$ and $d\mu/d\nu$ measured in our bilayer graphene device at $B=\SI{13.8}{T}$.  
In the $N=0$ Landau level, incompressible spikes are observed at fillings corresponding to the two- and four-flux `Jain' sequence\cite{jain_composite-fermion_1989}, with denominators as high as $17$. 
In the $N=1$ orbital, a different hierarchy is observed, including a prominent state at $\nu+1=1/2$ along with states at $8/17$ and $7/13$ filling.  This sequence is consistent with a Pfaffian state at half filling and abelian `daughter' states built from its elementary excitations\cite{levin_collective_2009,zibrov_tunable_2017}. Additional peaks are observed at fillings consistent with the four-flux Jain sequence, at $3/5$ and $2/5$, and finally several weaker states at $5/11$, $6/11$ and $5/9$ which were not previously reported.  Away from these incompressible fillings, the compressibility is negative throughout the partially filled Landau level\cite{eisenstein_negative_1992}. Additional negative compressibility is observed near the incompressible states, associated with the formation of Wigner crystals of fractionally charged quasiparticles at low quasiparticle density.

\begin{figure}[ht]
    \centering
    \includegraphics[width = 90mm]{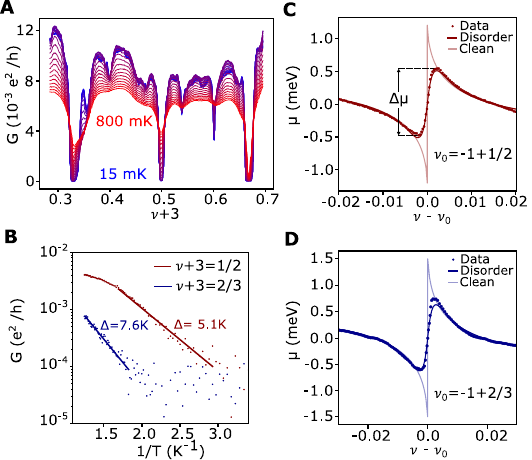}
\caption{\textbf{Comparison of activation and thermodynamic gap sin a partially filled $N=1$ Landau level.}
\textbf{(A)} Two terminal conductance measured in a Corbino geometry as a  function of filling factor at $B=\SI{12}{T}$ for different temperatures. The temperature spacing is 5mK. \textbf{(B)} Activation gap from the Arrhenius fit for $\nu=-3+1/2$ (red) and $\nu=-3+2/3$ (blue). \textbf{(C)}  
Chemical potential measurement near $\nu_0=-1+1/2$ (red dots) at $B=\SI{13.8}{T}$. Theory fit using the Wigner crystal model in the clean limit (light red line) and in the disordered limit (red line). \textbf{(D)} Chemical potential measurement near $\nu_0=-1+2/3$ (blue dots) at $B=\SI{13.8}{T}$. Theory fit using the Wigner crystal model in the clean limit (light blue line) and in the disordered limit (blue line).}
    \label{fig:2}
\end{figure}
Fig. 2A shows the two terminal conductance ($G$) measured at $B=\SI{12}{T}$ in a second sample consisting of a dual gated bilayer with Corbino-like geometry (see supplementary).  Measurements are taken at $B=\SI{12}{T}$ in a partially filled N=1 Landau level corresponding to filling factors  $0.25\lesssim \nu+3<0.75$ (see supplementary information). 
The three most prominent FQH states, at $\nu+3=1/3, 1/2$, and $2/3$, all show vanishing conductance at the lowest temperatures. Fig. 2B shows the minimal conductance for $\nu+3=1/2$ and 2/3 as a function of temperature, along with fits to an Arrhenius law $G \propto n_{\mathrm{qp}} \propto e^{-\Delta_{\mathrm{qp}}/2k_BT}$. 
For the $1/2$ state, the activation gap is found to be $ \Delta_{\mathrm{qp}} ^{\mathrm{act}}=\SI{5.1\pm0.2}{K}$ at $ B=\SI{12}{T}$, considerably larger than previous measurements in graphene\cite{zibrov_tunable_2017,li_even_2017} or other two-dimensional electron systems\cite{chung_ultra-high-quality_2021,watson_impact_2015,falson_even-denominator_2015,kumar_nonconventional_2010,shi_odd-_2020}.  

We may compare the result for the activation gap with a numerical calculation that accounts for the microscopic details of bilayer graphene, accomplished using the density matrix renormalization group (DMRG)~\cite{zaletel_topological_2013, mong_fibonacci_2017}. 
Following Ref.~\cite{zibrov_tunable_2017}, these calculations are conducted on an infinite cylinder within a 4-band model of BLG and account for mixing between the $N=0$ and $1$ Landau levels, screening from the gates, and---crucially--screening due to inter-Landau level transitions, which is treated within the random phase approximation (see supplementary information). 
We obtain a charge gap $\Delta_{\mathrm{qp}} ^{\mathrm{DMRG}}=0.011 E_C$, where the Coulomb energy scale $E_C$ depends on both the magnetic field and the dielectric constant for hBN, which we take as $\epsilon_{\mathrm{hBN}} = \sqrt{\epsilon_{xy} \epsilon_z} = 4.5$\cite{yang_experimental_2021}.  
The calculated gap is $\SI{5.6}{K}$ at $\SI{12}{T}$, within 10\% of the experimental value.

The jump in chemical potential at fractional filling, $\Delta \mu$, provides an alternative measurement of the FQH energy gaps, as shown in Figs. \ref{fig:2}C-D measured at $B=\SI{13.8}{T}$.  
In the clean limit, $\Delta \mu$  corresponds to the energy cost of adding a whole electron to the gapped system, and is expected to be $e/e^*$ times larger than the quasiparticle gap, where $e^*$ is the quasiparticle charge.  
At $\nu=-1/2$, where $e/e^*=4$, the quasiparticle gap $\Delta_{\mathrm{qp}}^{\mathrm{\mu}}=\Delta \mu/4=\SI{2.9}{K}$ implied by the measured thermodynamic is significantly smaller than $\Delta_{\mathrm{qp}} ^{\mathrm{act}} \approx \SI{5.1}{K}$, even before accounting for the small difference in $B$ between Figs. \ref{fig:2}A and C.  
A similar discrepancy is seen at $\nu+1=2/3$, where $\Delta_{\mathrm{qp}} ^{\mathrm{act}}=\SI{7.6\pm0.5}{K}$  but the quasiparticle gap from thermodynamic measurements is $\rm \Delta_{qp}^\mu=\SI{5.2}{K}$.  For comparison, at $\nu=1+2/3$ $\Delta_{\mathrm{qp}} ^{\mathrm{DMRG}}=\SI{11.7}{K}$.  

We attribute the discrepancy to the contrasting role of disorder on the thermodynamic and activation gaps.
In the simplest model for activated transport\cite{polyakov_universal_1995} disorder does not affect the activation gap (though in more involved models the activation gap becomes sensitive to the spatial correlations of the disorder potential\cite{dambrumenil_model_2011, nuebler_density_2010}).  
The thermodynamic gap, on the other hand, is sensitive to the presence of disorder-induced localized states which lead directly to a finite compressibility. 
To assess this hypothesis, we compare our data against a phenomenological model for $\mu(\nu)$ that accounts for both the disorder and quasiparticle interactions.  
Our model assumes that the compressible states adjacent to the incompressible FQH states are Wigner crystals of fractionally charged quasiparticles\cite{ezawa_fractional_1992,eisenstein_negative_1992}. 
As a starting point, we compute the energy density $\mathcal{E}( \nu)$ of this pristine Wigner crystal under the assumption that the fractional point charges $e^\ast$ form a triangular lattice and interact through an effective Coulomb potential which accounts for screening from the gates as well as the dielectric response of the parent gapped state. 
In the disorder-free limit, we obtain theoretical $\mu(\nu)$ curves in which an infinitely-sharp jump of $\Delta \mu = \frac{e}{e^*} \Delta_{\mathrm{qp}}$ is flanked by the negative compressibility of the screened Wigner crystal (see supplementary information). 
As shown in Figs. \ref{fig:2}C-D, we find this disorder-free model provides a good fit to the data at moderate quasiparticle densities, where the compressibility is strongly negative.  

To account for disorder, we make the assumption  that the disorder potential varies  slowly in comparison with both the inter-quasiparticle distance and the distance to the gates. 
As described in the supplementary material, this allows us to make a local density approximation; $\mu(\nu)$ can then be solved for explicitly given the interaction energy density $\mathcal{E}(\nu)$ and the disorder distribution $P[V_D]$, which we assume to be a Gaussian of width $\Gamma$.
We note that these assumptions may not be correct. For example, it will not be the case if the disorder arises from dilute Poisson-distributed charge impurities in the hBN. 
Nevertheless, it results in a tractable model that accounts for the competition between disorder and interactions.

Fits to this model are shown  in Fig.\ref{fig:2}C-D near $\nu = -1 + 1/2$ and $\nu=-1+2/3$. 
The fit is parameterized by the quasiparticle gap $\Delta^{\rm fit}_{\mathrm{qp}}$, a phenomenological parameter $\chi$ which accounts for the dielectric response of the parent state, and the disorder broadening $\Gamma$ (see supplementary information). We find quantitative agreement between the Wigner crystal model and experiment, providing strong evidence for a Wigner crystal of fractionalized quasiparticles.  
From the fit we infer $\Delta^{\mathrm{fit}}_{\mathrm{qp}}= \SI{7}{K}$ for the 1/2 state, within 20\% of $\Delta_{\mathrm{qp}}^{\mathrm{DMRG}} = \SI{6.0}{K}$. 
The same analysis for the $\nu_0 =-1 + 2/3$ gives $\Delta^{\mathrm{fit}}_{\mathrm{qp}}= \SI{11.6}{K}$, again within 20\% of the $\Delta_{\mathrm{qp}}^{\mathrm{DMRG}} = \SI{11.7}{K}$.
For both fillings, we find $\Gamma = \SI{1.0 \pm 0.5}{meV}$, consistent with previous estimates for the Landau level broadening \cite{polshyn_quantitative_2018,zeng_high-quality_2019}. The comparison between experimental and theoretical gaps is summarized in Table~\ref{tab:main_gap}.  

\begin{table}[h]
    \renewcommand*{\arraystretch}{1.3}
        \newcolumntype{C}{>{\centering\arraybackslash}X}
    \centering
\begin{tabularx}{0.48\textwidth}{|C|C|C|C|C||C|}
\hline  
Filling $\nu + 1$     & B &$\Delta_{\mathrm{qp}}^{\mathrm{\mathrm{act}}}  $ & $\Delta_{\mathrm{qp}}^{\mathrm{\mu}}$ & $\Delta_{\mathrm{qp}}^{\mathrm{fit}}$ &$\Delta_{\mathrm{qp}}^{\mathrm{DMRG}}$   \\ \hline
\multirow{2}{*}{$\displaystyle \frac12 $}& \SI{12}{T}   &\SI{5.1}{K}& $-$           & $-$& \SI{5.6}{K}\ \\\cline{2-6}
& \SI{13.8}{T} &$-$          & \SI{2.9}{K} & \SI{7.0}{K} & \SI{6.0}{K}\\ \hline 
\multirow{2}{*}{$\displaystyle \frac23$}& \SI{12}{T}   &\SI{7.6}{K}& $-$           & $-$& \SI{10.8}{K}\\\cline{2-6}
& \SI{13.8}{T} &$-$          & \SI{5.2}{K} & \SI{11.6}{K}& \SI{11.7}{K}\\\hline 
    \end{tabularx}
    \caption{Comparison of the quasiparticle gaps at $1/2$ and $2/3$ filling in the $N = 1$ Landau level as determined by DMRG calculations $\Delta_{\mathrm{qp}}^{\mathrm{DMRG}}$, thermally activated transport $\Delta_{\mathrm{qp}}^{\mathrm{act}}$, the chemical potential jump $\Delta_{\mathrm{qp}}^{\mu}$, and from the fit to the Wigner crystal model $\Delta_{\mathrm{qp}}^{\mathrm{fit}}$. }
    \label{tab:main_gap}
\end{table}
Given the rather large discrepancies between experiment and numerics in GaAs\cite{ma_fractional_2022}---particularly at half filling---the level of agreement we find for both activated and thermodynamic gaps with numerical modeling is encouraging. We note that several sources may account for the remaining quantitative discrepancies in our work. These including differences in inter-Landau level screening strength at $\nu \sim -3$ relative to $\nu \sim -1$\cite{shizuya_electromagnetic_2007}, as well as possible spin textures in the excitation spectrum, which can lower the activation gap but are not accounted for in our modeling.  For $\rm \Delta_{qp}^{fit}$, moreover, the phenomenological nature of our model for disorder may not capture the microscopic physics at a quantitative level.  

\begin{figure}[ht]
    \centering
   \includegraphics[width = 88mm]{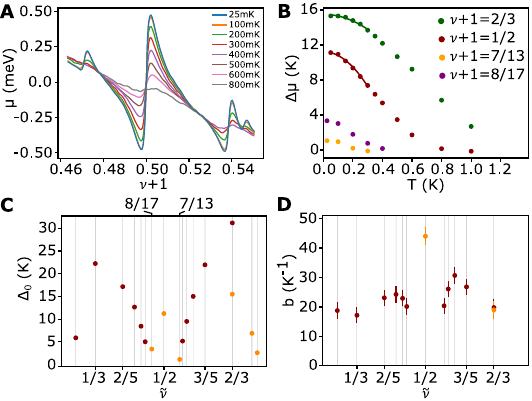}
    \caption{\textbf{Temperature dependent $\mu$ near fractional filling at B=13.8T.}
    \textbf{(A)} Chemical potential near half filling of an $N=1$ Landau level at several different temperatures. 
    \textbf{(B)} Chemical potential jump across the incompressible states as a function of temperature for different filling factors in an $N=1$  LL (dots). The solid lines are a low temperature fit, $\Delta \mu(T) =\Delta_0-bT^2$.  
    \textbf{(C)} Chemical potential jump $\Delta_0$ extracted from the  fit for different fractional states in the N=0 ($\tilde{\nu}=\nu$, red dots) and  (N=0 orbital) and $N=1$ ($\tilde{\nu}=\nu+1$, orange dots) orbital Landau levels. 
    \textbf{(D)} Temperature decay parameter $b$  extracted from the  fit same states. }
    \label{fig:3}
\end{figure}

Fig. \ref{fig:3}A shows the $\mu$ measured at different temperatures near the $\nu+1=1/2$ gap. We focus on the strong temperature dependence of $\Delta \mu$, plotted for several incompressible filling factors in Fig. \ref{fig:3}B (see also the supplementary information).  
We fit the low temperature limit of $\Delta \mu(T)$ using the Sommerfeld expansion $\Delta \mu(T)=\Delta_0-b T^2 + \cdots$, which is justified so long as the quasiparticles experience short-range repulsion. 
The fitted values $\Delta_0$ and $b$ are reported in Figs. \ref{fig:3}C and D, respectively.

Notably, the $\nu=-1+1/2$ state shows anomalously strong temperature dependence, manifesting as a large value of the $b$ parameter. 
According to the Maxwell relation $ |\frac{d\mu}{dT}|_{n}=-|\frac{ds}{dn}|_{T}$, this suggests an anomalous contribution to the entropy in the dilute quasiparticle limit. Anomalous entropy is expected in the vicinity of non-Abelian states\cite{cooper_observable_2009} owing to the topological degeneracy of a dilute gas of nonabelian anyons.  However, this contribution is considerably smaller than the anomalous entropy we observe.  
To see this, we assume the quasiparticles are Ising anyons, endowing the ground state with an anomalous entropy of $S_{\mathrm{topo}} = k_B N_{\mathrm{qp}} \ln(\sqrt{2})$
\cite{cooper_observable_2009}. 
Accounting for this contribution adds a linear-in-$T$ term to the low-temperature expression, $\Delta\mu(T)=\Delta_0 -4 \ln(2)T -bT^2 + \cdots$. Refitting the $1/2$ state to account for this modification reduces the best-fit $b$ to \SI{36 \pm 3}{K^{-1}}--still larger than $b$ for all odd-denominator states. 

This analysis implies that the anomalous entropy near $\nu=1/2$---at least at the filling factors corresponding to the extrema in $\mu$---does not arise solely from the topological degeneracy. Notably, these extrema occur at a density of quasiparticles where the average inter-quasiparticle distance is larger than the distance to the gate.  Disorder is expected to dominate this regime, as inter-quasiparticle interactions are screened.  Crudely, if disorder is more important than the long-range Coulomb interaction, we expect $b \propto (e / e^\ast)^2 / \Gamma$, where $\Gamma$ is the strength of the disorder. However, determining the prefactor requires understanding the thermodynamics of a Coulomb glass of fractionalized particles in an unknown disorder distribution, a challenge we leave to future work.  

In closing, we note that a related manuscript reports scanning tunneling microscopy to study the same bilayer graphene FQH states studied here\footnote{A. Yazdani, Private communication}. 
In that work, the gate voltage $\delta V_g$ over which the FQH gaps appear provides a local measurement of the thermodynamic gap.  Those authors find $4 \Delta^{\textrm{STM}}_{\rm qp} = \SI{30}{K}$ for the 1/2 state at $B=\SI{14}{T}$.  This result is consistent with the intrinsic gap inferred from our WC model, $4 \Delta^{\textrm{WC}}_{\rm qp} \sim \SI{28}{K}$, as expected for a local measurement that probes the chemical potential at length scales smaller than the disorder correlation length. 
The large intrinsic gaps manifesting across several experimental techniques show that bilayer graphene is an ideal platform to explore the intrinsic physics of nonabelian anyons in the solid state.

\begin{acknowledgments}
The authors would like to acknowledge  discussions with A. Stern, and A. Yazdani for a related collaboration and sharing unpublished results, and Evgeny Redekop for providing the device image shown in Fig. 1A.   
Experimental work at UCSB was primarily supported by the Office of Naval Research under award N00014-23-1-2066 to AFY.  
AFY acknowledges additional support by  Gordon and Betty Moore Foundation EPIQS program under award GBMF9471.
MZ and TW were supported by the U.S. Department of Energy, Office of Science, Office of Basic Energy Sciences, Materials Sciences and Engineering Division, under Contract No. DE-AC02-05CH11231, within the van der Waals Heterostructures Program (KCWF16).
RM is supported by the National Science Foundation under Award No. DMR-1848336. 
KW and TT acknowledge support from the Elemental Strategy Initiative conducted by the MEXT, Japan (Grant Number JPMXP0112101001) and JSPSKAKENHI (Grant Numbers 19H05790, 20H00354 and 21H05233). This research used the Lawrencium computational cluster provided by the Lawrence Berkeley National Laboratory (Supported by the U.S. Department of Energy, Office of Basic Energy Sciences under Contract No. DE-AC02-05CH11231)
\end{acknowledgments}

\bibliographystyle{apsrev4-1}

%

\clearpage
\newpage
\pagebreak

\onecolumngrid

\begin{center}
\textbf{\large Supplementary information }\\[5pt]
\end{center}

\setcounter{equation}{0}
\setcounter{figure}{0}
\setcounter{table}{0}
\setcounter{page}{1}
\setcounter{section}{0}
\makeatletter
\renewcommand{\theequation}{S\arabic{equation}}
\renewcommand{\thefigure}{S\arabic{figure}}
\renewcommand{\thetable}{S\arabic{table}}
\renewcommand{\thepage}{\arabic{page}}

\section{Experimental methods}
\subsection{Sample preparation}

\begin{figure*}[ht]
    \centering
    \includegraphics[width = 180mm]{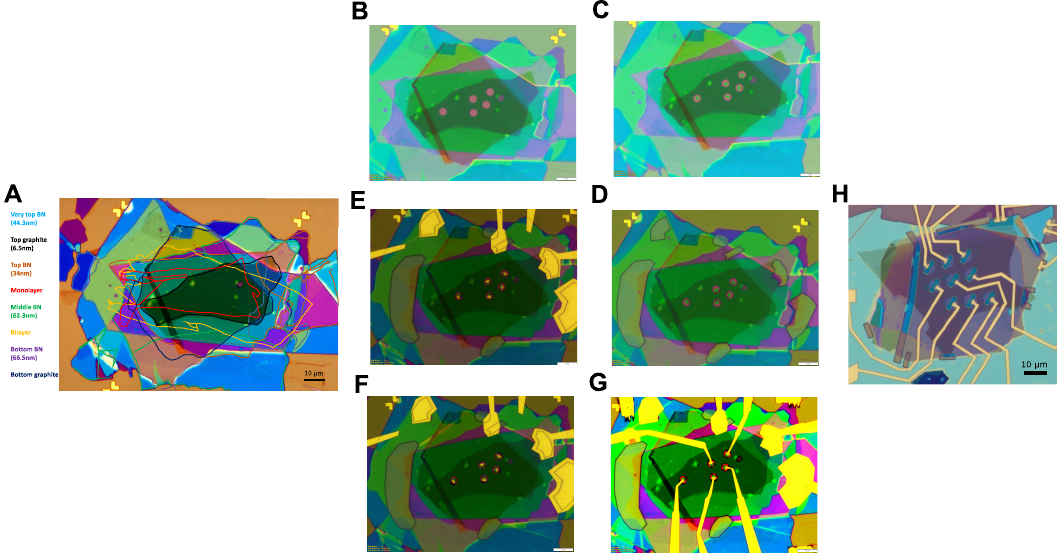}
    \caption{\textbf{Sample fabrication chronological steps.}
    \textbf{(A)} Optical microscope image of the final heterostructures with the layer borders highlighted with colors \textbf{(B)} Holes are defined in the top gates. \textbf{(C)} Smaller hole in a Pac-Man shape to the monolayer are etched within the top gate holes. \textbf{(D)} Contact trenches are defined together with openings to contact all the layers. \textbf{(E)} Metal deposition to make contact to the different layers. \textbf{(F)} Overdose PMMA is used to define bridges on top of the exposed top gate edges. \textbf{(G)} Final image of the chemical potential device. \textbf{(H)} Final image of the Corbino transport device.}
    \label{fig:fab}
\end{figure*}

The van der Waals heterostructure is assembled using a polycarbonate based dry stacking technique. An image of the final heterostructure is shown in Fig S1A together with all the different hBN thicknesses, measured by atomic force microscopy. Two subsequent lithography steps are performed: first, we define holes in the top gate (Fig. \ref{fig:fab}B); then we define smaller holes in a `Pac-Man' shape within the top gate holes to expose the graphene edge (Fig. \ref{fig:fab}C). The etching is realized with a RIE using CHF3/O2 gas. Another etch is done to define trenches in the monolayer, so that the quantum Hall edge states are connected to the bulk of the device (Fig. \ref{fig:fab}D). Metal deposition of Cr/Pd/Au (3nm/15nm/120nm) makes electrical contact to the bilayer graphene, monolayer graphene, and  gate  layers (Fig. \ref{fig:fab}E). In order to connect the isolated graphene contacts without shorting them to the exposed top gate edges, we use overdosed PMMA bridges (Fig. \ref{fig:fab}F). Finally the graphene contacts are connected to macroscopic leads.  The finished device shown in Fig. \ref{fig:fab}G. 

For the Corbino transport measurements, the sample consists of a dual-graphite gated, hBN encapsulated Bernal bilayer graphene layer, shown in Fig. \ref{fig:fab}H. The top hBN thickness is 48.9nm and the bottom hBN thickness is 37.2nm. A transport phase diagram from this sample is shown in Fig. \ref{fig:phaseDiag}B. 

\subsection{Chemical potential measurements}

\begin{figure*}[ht]
    \centering
    \includegraphics[width = \textwidth]{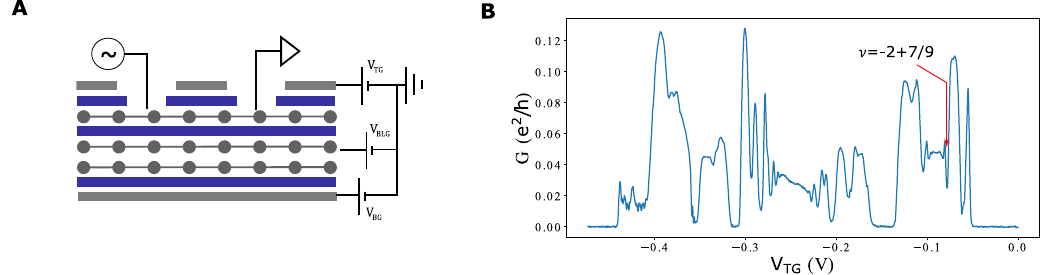}
    \caption{\textbf{Measurement principle.}
    \textbf{(A)} Device schematic showing the dual graphite gated sample with a graphene monolayer detector capacitively coupled to the bilayer graphene sample of interest. \textbf{(B)} Monolayer detector conductance as a function of top gate voltage at B=13.5T. The four flux fractional state conductance minima $\nu=-2+7/9$, highlighted with the red arrow, is used as a sharp detector of the chemical potential change in the bilayer graphene sample.
    }
    \label{fig:schem}
\end{figure*}

A schematic of the chemical potential measurement scheme is shown in Fig. \ref{fig:schem}A. A graphene monolayer transducer is placed on top of the bilayer sample in a dual graphite gated device. We tune the monolayer density to a sharp conductance minima defined at fractional filling $\rm \nu=-2+7/9$ with the top gate voltage, as shown in Fig. \ref{fig:schem}B. While the bilayer is grounded, its density is adjusted using the back gate voltage. The chemical potential shift of the bilayer can then be detected via the  shift of the monolayer conductance minima on the top gate axis.

We model our system as a four plate capacitor model which accounts for the chemical potential of the bilayer and of the monolayer detector.  The system electrostatics are described by the relation:  

 \begin{equation}
     \hat C (\vec V+\vec \mu /e)=-e\vec n,
     \label{electrostatics}
 \end{equation}

 Where $\hat C$ is the geometric capacitance matrix, $\vec V$ are the applied voltages, $\vec n$ is the vector of charge carrier densities, and $\vec \mu$ is the chemical potential of the layers which we take to be fixed for the top and bottom gates but is a (density dependent) quantity for the sample and detector layers. 
 \begin{equation}
     \hat C=
  \left( {\begin{array}{ccccc}
    -c_{td} & c_{td} & 0 & 0 \\
    c_{td} & -c_{td}-c_{ds} & c_{ds} & 0 \\
    0 & c_{ds} & -c_{ds}-c_{sb} & c_{sb}\\
    0 & 0 & c_{sb} & -c_{sb}\\
  \end{array} } \right), \quad 
       \vec V=
  \left( {\begin{array}{ccccc}
    V_{t} \\
    V_{d} \\
    V_{s} \\
    V_{b} \\
  \end{array} } \right), \quad 
       \vec \mu=
  \left( {\begin{array}{ccccc}
    0 \\
    \mu_{d}(n_d) \\
    \mu_{s}(n_s) \\
    0 \\
  \end{array} } \right),
  \quad 
       \vec n=
  \left( {\begin{array}{ccccc}
    n_{t} \\
    n_{d} \\
    n_{s} \\
    n_{b} \\
  \end{array} } \right)
 \end{equation}

Here, $c_{ij}$ is the geometric capacitance between the layers $i$ and $j$, with $i,j=t,d,s,b$, indicating the top gate, detector layer,
sample layer, and bottom gate, respectively. Note that $c_{ij}=c_{ji}$.

In our experiment, we vary the bottom gate voltage by $\Delta V_{b}$, while leaving the sample at constant voltage so that $\Delta V_s=0$.  We then find the top gate voltage $\Delta V_{t}$ such that the  detector density $\delta n_t=0$, which in turn implies $\Delta \mu_t=0$.  Under these conditions, we have the relation 
  \begin{equation}
     \Delta V + \Delta \mu /e=
  \left( {\begin{array}{ccccc}
    \Delta V_{t} \\
    0 \\
    \Delta \mu_{s}/e \\
    \Delta V_{b} \\
  \end{array} } \right).
 \end{equation}

Using Eq. \eqref{electrostatics}, the second row of the matrix gives

  \begin{align}
    c_{td} \Delta V_t  + c_{ds} \Delta \mu_s /e&=-e \Delta n_d=0\\
     \Delta \mu_s &=-e \Delta V_t \frac{c_{td}}{c_{ds}} 
 \end{align}

Giving the bilayer chemical potential shift from the measured shift on the top gate.  
The third row of Eq. \eqref{electrostatics} gives an expression for the \textit{density} of the sample layer $\Delta n_s$, 
 
  \begin{equation}
     \Delta n_s =  -(c_{sd}+c_{sb}) \Delta \mu_s /e + c_{sb} \Delta V_b
 \end{equation}

\begin{figure*}[ht]
    \centering
    \includegraphics[width = 180mm]{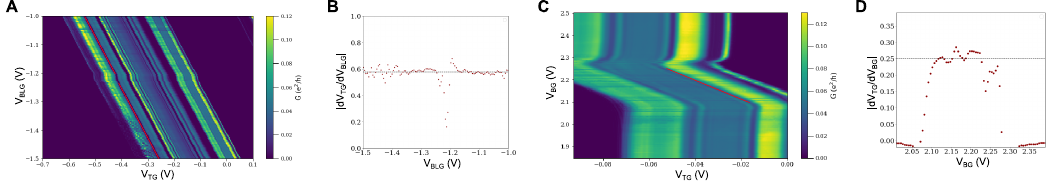}
    \caption{\textbf{Capacitance ratio.}
    \textbf{(A)} Monolayer conductance map as a function of bilayer gate and top gate. \textbf{(B)} Derivative of the top gate shift as a function of bilayer gate. The black dashed line shows the value of the capacitance ratio $c_{ds}/c_{td}=0.58$. \textbf{(C)} Monolayer conductance map as a function of back gate and top gate across $\nu=4$. \textbf{(D)} Derivative of the top gate shift as a function of back gate. The black dashed line shows the value of the capacitance ratio $\frac{c_{sb}c_{ds}}{(c_{sb}+c_{ds})c_{td}}=0.26$.
    }
    \label{fig:ratio}
\end{figure*}

Using the relations above requires accurate knowledge of the geometric capacitances.  Particularly the ratios $\rm \frac{c_{d,t}}{c_{d,s}}$ and $\rm \frac{c_{s,d}}{c_{s, b}}$. This is done by tracking the conductance minima as a function of various gate voltage as shown in Fig. \ref{fig:ratio}. In these experiments, a fixed density feature in the detector layer is tracked through a region where the sample layer is incompressible (for example, in the $\nu=4$ integer quantum Hall gap) or in the middle of the compressible Landau level.  This allows the relevant capacitance ratios to be measured directly with high accuracy.

\begin{figure*}[ht]
    \centering
    \includegraphics[width = 150mm]{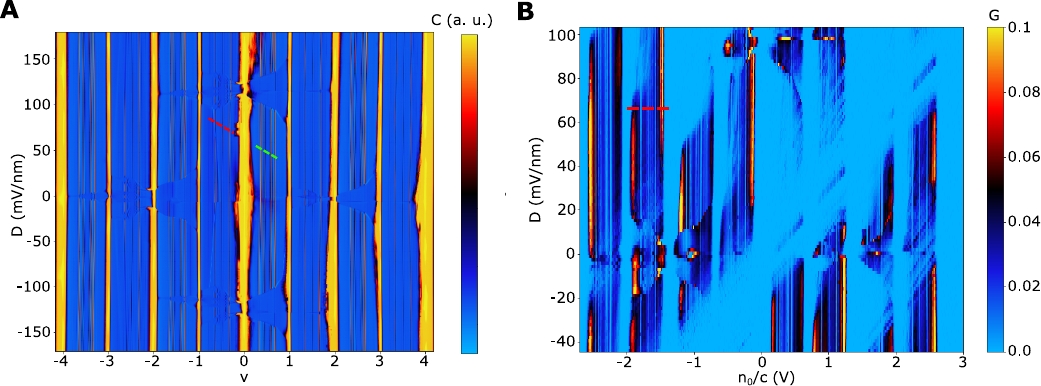}
    \caption{ \textbf{Capacitance and transport phase diagram of bilayer graphene.} \textbf{(A)} Capacitance map from Device B of  Ref. \cite{zibrov_tunable_2017} at B=14T. The green (N=0 orbital) and red (N=1 orbital) dashed lines correspond to trajectories shown in the main text Figs 1C and 1D, respectively. \textbf{(B)} Conductance map from the bilayer Corbino device at B=12T. The red dashed line correspond to the line cut where the main text data shown in Fig 2A has been measured.
     }
    \label{fig:phaseDiag}
\end{figure*}

A final electrostatic consideration for bilayer graphene is the displacement field $D$, which modifies the precise nature of the filled Landau level orbitals, favoring valley polarization at large $D$.  We obtain the displacement field from the relation $D=\frac{1}{2\epsilon_0} (c_{sd} (V_d-V_s)-c_{sb} (V_b-V_s))$. 
While we do not map the entire phase diagram of our bilayer sample as a function of $D$, this phase diagram is highly reproducible across devices\cite{zibrov_tunable_2017}.  In Fig. \ref{fig:phaseDiag}A, we show the trajectories in this phase diagram acquired on another device (Device B of reference \cite{zibrov_tunable_2017}) corresponding to the data shown in the main text. Also, the transport phase diagram showing where the activation gaps have been measured is shown in Fig. \ref{fig:phaseDiag}B.

\newpage 

\section{Computational methods}

\subsection{DMRG calculation of the charge gaps}

To find the charge gaps at the incompressible fractional quantum Hall (FQH) states in bilayer graphene (BLG), we first use infinite density matrix renormalization group (iDMRG) on a cylinder to obtain the ground state at the corresponding fractional fillings $\nu$, following the formulation developed in Ref.~\onlinecite{zibrov_tunable_2017}. In the DMRG calculation, we always assume full isospin polarization, which is justified experimentally in Ref.~\onlinecite{zibrov_tunable_2017,hunt_direct_2017}. Compared to the conventional lowest Landau level (LLL) in GaAs, the ``zeroth'' Landau level (ZLL) in bilayer graphene has an additional $N=1$ orbital, consisting of a mixture of conventional $n=0$ and $n=1$ LLs. We keep both the $N=0$ and $N = 1$ orbitals within the ZLL and write the ZLL-projected density operator as 
\begin{equation}
    n_{\mathrm{ZLL}}(q)=\sum_{N, N^{\prime} = 0}^1 \mathcal{F}_{N, N^{\prime}}(q) \bar{\rho}_{N N^{\prime}}(q)
\end{equation}
where $\bar{\rho}_{N N^{\prime}}(q)$ is the usual density operator projected into $N, N^{\prime}$ orbitals and $\mathcal{F}$ are the BLG ``form factors'' defined in Ref.~\onlinecite{zibrov_tunable_2017}. The remaining $|N| \geq 2$ LLs are separated from the ZLL by at least the cyclotron gap $\hbar \omega_c$, whose main effect is to screen the interaction via inter-LL virtual excitations which can be captured by the effective interaction discussed below.

The bare Coulomb interaction is screened by the encapsulating hBN, the graphite gates, and inter-Landau level (LL) transition. Screening from the $\mathrm{hBN}$ and the graphite gates can be captured by
\begin{equation} \label{eq:V0q}
    V^0(\mathbf{q})=E_C \ell_B \frac{4 \pi \sinh \left(\beta d_t|\mathbf{q}|\right) \sinh \left(\beta d_b|\mathbf{q}|\right)}{\sinh \left(\beta\left(d_t+d_b\right)|\mathbf{q}|\right)|\mathbf{q}|}
\end{equation}
where $E_C=e^2 / 4 \pi \epsilon_0 \epsilon_{\mathrm{hBN}} \ell_B$ is the Coulomb scale, $d_t$ ($d_b$) are the distance from the sample to the the top (bottom) graphite gate. Here $\epsilon_{\mathrm{hBN}}=\sqrt{\epsilon_{\perp} \epsilon_{\|}}$ and $\beta=\sqrt{\epsilon_{\|} / \epsilon_{\perp}}$ are defined with the out-of-plane (in-plane) dielectric constant $\epsilon_{\perp}$ ($\epsilon_{\|}$) of hBN. In all DMRG calculations we stick to $d_t = d_b = \SI{60}{nm}$ mimicking the device geometry. The precise dielectric constants of hBN have been subject to debate, but we pick $\epsilon_{\perp} = 3.15$ and $\epsilon_{\|} = 6.6$ from Ref.~\onlinecite{geick_normal_1966} and our own measurement.

\begin{figure*}[htbp]
    \centering
    \includegraphics[width = 60mm]{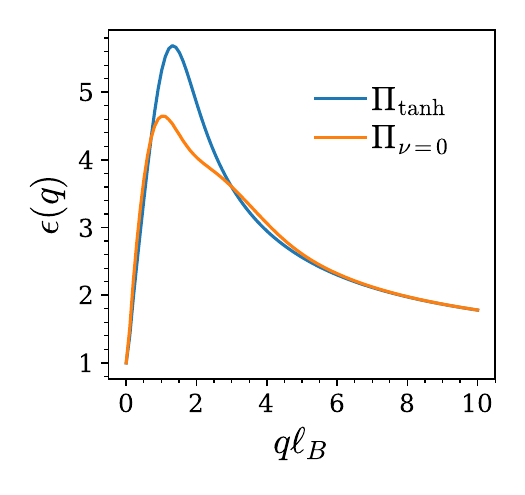}
    \caption{Dielectric function in BLG at $B = \SI{14}{T}$ with $V^0(\mathrm{q})$ shown in Eqn.~\ref{eq:V0q}. We show the one obtained with the calculated polarization function $\Pi_{\nu}(\mathrm{q})$ in Eqn~\ref{eq:Pi} with $U = \SI{8}{meV}$ and $\nu = 0$ and the one obatined with the phenomenological polarization function $\Pi_{\mathrm{tanh}}(\mathrm{q})$ in Eqn.~\ref{eq:tanh}.}
    \label{fig:pi}
\end{figure*}

The static dielectric response due to inter-LL virtual excitations can be obtained within the random phase approximation (RPA),
\begin{equation}
    V_{\mathrm{RPA}}(\mathbf{q}) = \frac{V_0(\mathbf{q})}{1-V_0(\mathbf{q}) \Pi_\nu(\mathbf{q}, \omega=0)}
    \label{eq:RPA}
\end{equation}
where the polarizability $\Pi_\nu(\mathbf{q})$ sums over all inter-LL transitions $m \to n$ except for $0 \leftrightarrow 1$.
\begin{equation} \label{eq:Pi}
    \Pi_\nu(\mathbf{q})=\sum_{-\Lambda<m, n<\Lambda} \nu_m\left(1-\nu_n\right) \Pi_{m, n}(\mathbf{q})
\end{equation}
 where $m, n$ label LLs, $\nu_m$ is the filling of $m$th LL, and $\Lambda \gg 1$ is a high energy cutoff. For simplicity, we follow Ref.~\onlinecite{misumi_electromagnetic_2008} by calculating $\Pi_{m, n}(\mathrm{q})$ within a two-band model of BLG assuming only $\gamma_0, \gamma_1 \neq 0$. We use inter-layer bias $U = \SI{8}{meV}$, with orbital filling in the order of $\ket{K' \downarrow 0},\ket{K' \downarrow 1},\ket{K \downarrow 0},\ket{K \downarrow 1},\ket{K' \uparrow 0},\ket{K' \uparrow 1},\ket{K \uparrow 0},\ket{K \uparrow 1}$ at $B > \SI{12}{T}$ where $\ket{\tau,s,N}$ labels the valley, spin, and LL \cite{li_even_2017,hunt_direct_2017}. We note that any transition $0 \leftrightarrow 1$ within the ZLL is disregarded since the DMRG calculation automatically accounts for such transitions by keeping both the $N = 0$ and $N = 1$ orbitals. The sum in Eqn.~\ref{eq:Pi} converges slowly with the cutoff $\Lambda$, so we scale the cutoff from $\Lambda=110$ to $170$ and extrapolate to $\Lambda \to \infty$ \cite{yang_experimental_2021}. We show the resulting polarization function $\Pi_{\nu}(\mathbf{q})$ in Fig.~\ref{fig:pi}, which agree with the commonly used phenomenological form $\Pi_{\mathrm{tanh}}$ at large $q$ and small $q$ limit but deviate at intermediate $q$ \cite{hunt_direct_2017,papic_topological_2014,zibrov_tunable_2017}. The deviation, which depends on both the filling and displacement field, has a quantitative effect on the charge gaps.

After obtaining the FQH ground states in BLG, we use the ``defect'' DMRG method to compute the quasiparticle gaps \cite{zaletel_topological_2013}, which is implemented for multicomponent quantum Hall systems in Ref.~\onlinecite{mong_fibonacci_2017}. The key idea is to enforce a single anyonic excitation $a$ to exist in some large central region on an infinite cylinder using a ``topological boundary condition'' $[l, r]$, which means fixing the ground state sector to the left and right of the central region. We determine the energy of anyon $a$ by comparing the energy of this particular configuration to that of the vacuum. The quasiparticle gap is the energy required to create and separate a pair of lowest-lying quasiparticle and quasihole $\Delta=E_{e^*}+E_{-e^*}$. In Table~\ref{tab:root}, we show the lowest-lying quasiparticles and quasiholes at varioud FQH states and the corresponding ``topological boundary conditions'' required to trap them in ``defect'' DMRG.

\begin{table}[]
    \renewcommand*{\arraystretch}{1.5}
    \newcolumntype{C}{>{\centering\arraybackslash}X}
    \centering
    \begin{tabularx}{0.8\textwidth}{C|CC}
\hline \hline
                                                      & lowest-lying charge excitation       & topological boundary condition $[l,r]$                  \\ \hline
\multirow{2}{*}{Jain sequence $\nu = \frac{p}{2p+1}$} & $e^* = \frac{e}{2p+1}$ quasiparticle & {[}``$01 \cdots 010$'', ``$01 \cdots 001$''{]}  \\
                                                      & $e^* = -\frac{e}{2p+1}$ quasihole    & {[}``$01 \cdots 010$'', ``$10 01 \cdots 0$''{]} \\
\multirow{2}{*}{Pfaffian-like state $\nu = \frac12$}  & $e^* = \frac{e}{4}$ quasiparticle    & {[}``$0110$'', ``$1010$''{]}                    \\
                                                      & $e^* = -\frac{e}{4}$ quasihole   & {[}``$0110$'', ``$0101$''{]}                    \\ \hline \hline
\end{tabularx}
    \caption{Lowest lying charge excitations and the corresponding topological boundary conditions at various FQH states. Here we label different ground state sectors by their root configurations. ``$01 \cdots$'' means ``$01$'' repeating $p - 1$ times.}
    \label{tab:root}
\end{table}

\begin{figure*}[t]
    \centering
    \includegraphics[width = 160mm]{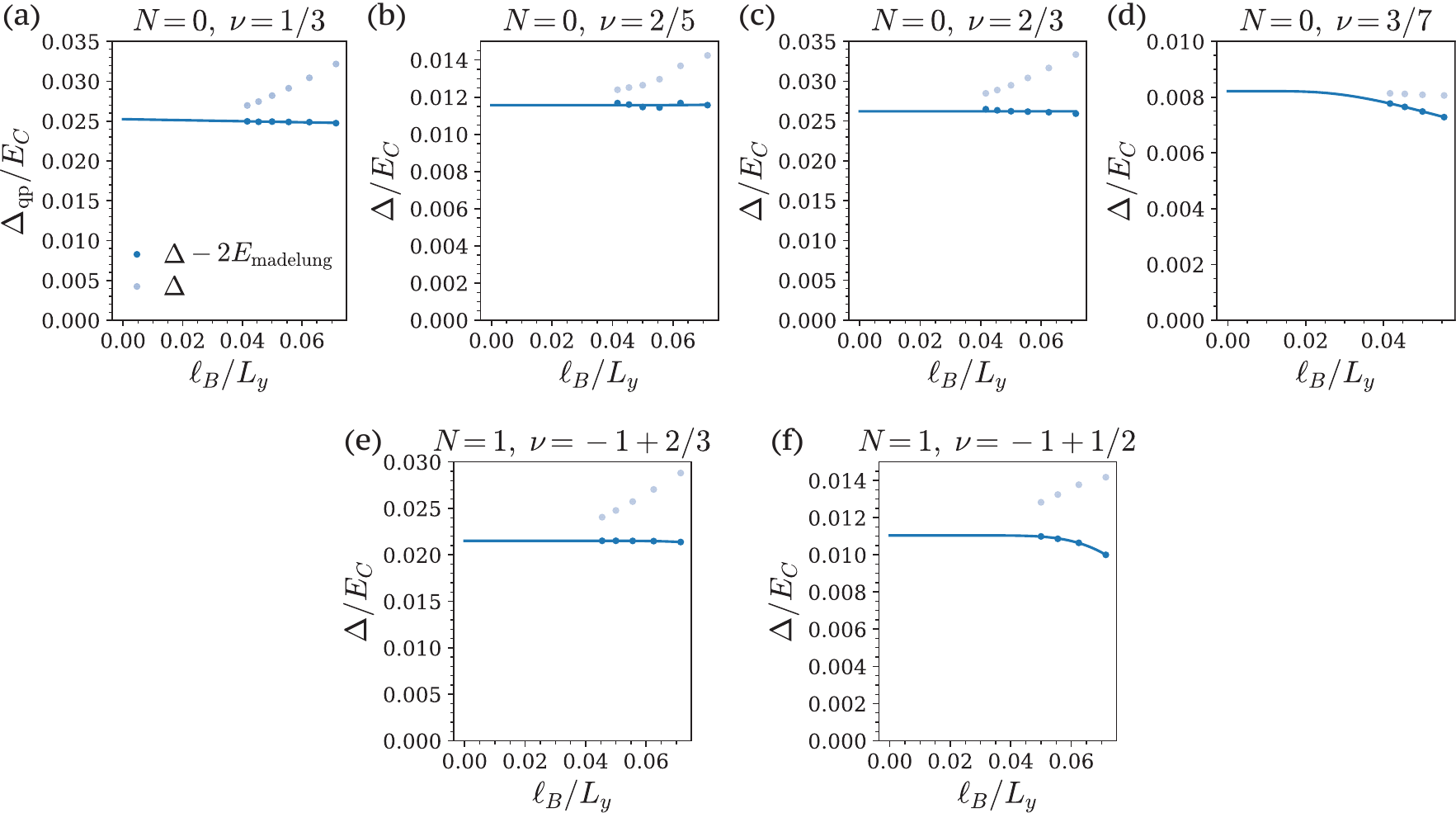}
    \caption{Quasiparticle gap at various FQH states in BLG calculated by DMRG. We subtract off $E_{\mathrm{madelung}}$ for the dark dots, which exhibit vanishing $L_y$ dependence except for at $\nu = 3/7$ and $\nu = 1/2$.}
    \label{fig:dmrg}
\end{figure*}

The accuracy of the calculation is primarily controlled by three parameters. The most important parameter is the cylinder circumference $L_y$ used in iDMRG. Due to the absence of local curvature on the cylinder, there are no geometric corrections typically seen in FQH numerics on the sphere \cite{mong_fibonacci_2017}. The Coulomb interaction is screened by the gates on two sides such that at sufficiently large $L_y$, we expect the interaction energy $E_{\mathrm{madelung}}$ between the anyon and its images around the cylinder to vanish exponentially. In practice, however, the gate distance $d = \SI{60}{\nm}$ is comparable to the accessible range of $12 \ell_B \leq L_y \leq 24 \ell_B$, so we still need to manually subtract off $E_{\mathrm{madelung}}$. For simplicity, we approximate the anyon on the cylinder by a point charge at $y = 0$, then $E_{\mathrm{madelung}}$ can be computed by
\begin{equation}
    E_{\mathrm{madelung}}(L_y) = \frac{(e^*)^2}2 \sum_{n \in \mathbb{Z}, n \neq 0} V_{\mathrm{RPA}}(nL_y)
\end{equation}
where $e^*$ is the charge of the anyon, and the mirror images locate at $y = n L$ with $n$ being a non-zero integer due to the periodic boundary condition along the $y$ axis. In the calculation of the charge gaps we consider a pair of quasiparticle and quasihole, each contributing to one copy of the interaction energy $E_{\mathrm{madelung}}$. In Fig.~\ref{fig:dmrg}, we show the quasiparticle gaps computed by DMRG. After subtracting of two copies of $E_{\mathrm{madelung}}$, the resulting gaps exhibit vanishing $L_y$ dependence, except for at $\nu = 3/7$ and $\nu = 1/2$ which requires a larger $L_y$ to stablize the ground states. At these two states, we perform an exponential extrapolation in $L_y$ by $\Delta\left(L_y\right)=\Delta+\delta e^{-L_y / L_{y 0}}$ and report $\Delta$.
 
Another obvious parameter is the bond dimension $\chi$ within DMRG, which determines the maximum allowed entanglement in the system. In practice, the charge gaps change negligibly with bond dimension $\chi \geq 1600$ up to the largest cylinder circumference $L_y = 24 \ell_B$ for all the FQH states we examine. A more subtle but equally important parameter is the number of orbitals $N_{\mathrm{defect}}$ in the central region in ``defect'' DMRG. The ``defect'' DMRG algorithm functions by introducing $N_{\mathrm{defect}}$ tensors in the middle of two infinite matrix product state (MPS) ground states on two sides, and then optimizing this configuration with a finite DMRG algorithm. It is crucial for the corresponding physical span $\frac{2 \pi}{L_y} \times N_{\mathrm{defect}}$ to exceed the anyon radius $\xi_a$, which is typically a few $\ell_B$. We stick to $N_{\mathrm{defect}} = 128$ which turns out to be sufficient up to the largest cylinder circumference $L_y = 24 \ell_B$ for all FQH states except for at $\nu = 1/2$ where Ising anyons have a much longer tail. At $\nu = 1/2$, we extrapolate the span $N_{\mathrm{defect}} = 64$ to $128$ and restrict ourselves to $L_y \leq 18 \ell_B$.
\\

\subsection{Quasiparticle Wigner crystal model}

We model the compressible states adjacent to the incompressible fractional quantum Hall states as Wigner
crystals of fractionally charged quasiparticles. We start by thinking of electron Wigner crystals, whose energy can be evaluated as
\begin{equation}
\label{eq:WCenergy}
    E_{\mathrm{WC}}=\sum_{i \neq j} \sum_{\left\{\bR_i\right\}} \frac{1}{2} \frac{\left\langle\Psi\left|V\left(\left|\bR_i-\bR_j\right|\right)\right| \Psi\right\rangle}{\langle\Psi \mid \Psi\rangle} \approx \frac{N_e}{2} \sum_{\bR_i \neq 0} V_H\left(\left|\bR_i\right|\right)
\end{equation}
where $N_e$ is the number of the electrons, $V(\bR)$ is the effective interaction in Eqn.~(\ref{eq:RPA}), and $\bR_i$ spans the triangular lattice of the Wigner crystal. Here we plug in the zeroth Landau level (ZLL) wavefunctions (which include both the $N=0, 1$ LLs)  and expand up to terms that contain two of the $\bR$'s, neglecting all Fock terms \cite{maki_static_1983}. 
There is a simple relation between $V_H(\bR)$ and $V(\bR)$ in Fourier space, i.e. $V_H(\bq) = V(\bq) \left |F(\bq) \right|^4$ with $F(\bq) = e^{- \bq^2 \ell_B^2/4}$ being the LLL form factor.

The classical treatment of the WC is valid as long as the electrons are sufficiently far away from each other $R \gg \ell_B$, then we can neglect quantum statistics, which always holds in the filling range of interest.

The relevant energy that determines the chemical potential in the experiment is the WC energy relative to the charging energy of a classical capacitor, 
\begin{equation} \label{eq:tanh}
    E_{\mathrm{int}} = E_{\mathrm{WC}} - \frac{e^2 N^2}{2 C_g}
\end{equation}
where $C_g$ is the geometric self capacitance of the bilayer graphene plate. Then the desired chemical potential $\mu(\nu)$ in the clean limit is given by $\mu_{\mathrm{int}}(\nu) = \partial_{\nu} ( E_{\mathrm{int}}/N_{\phi} )$ with $N_{\phi}$ being the number of fluxes.

The formulation discussed so far has primarily focused on Wigner crystals of electrons, which have predicted a chemical potential in excellent agreement with experimental results near integer quantum Hall states in monolayer graphene \cite{zibrov_tunable_2017}. However, our main focus is on Wigner crystals of quasiparticles. These fractional quasiparticles carry a charge of $e^*$ and possess their own Landau levels, where the magnetic length is effectively increased to $\tilde{\ell}_B = \sqrt{\hbar / e^* B}$. Within the composite-fermion picture, the quasiholes, such as those in the $\nu=\frac{1}{3}$ state, even exhibit identical wavefunctions to the lowest Landau level but with $\ell_B$ replaced by $\tilde{\ell}_B$. In the filling range of interest in this study, these quasiparticles are sufficiently far apart from each other that their interactions are expected to be be reasonably well approximated as those of electric monopoles, therefore ignoring details of the charge distribution. Therefore, a rough estimate of the chemical potential adjacent to the fractional quantum Hall states can be made by substituting $e$ with $e^*$ and $\ell_B$ with $\tilde{\ell}_B$. One additional complication comes from the screening from composite Fermion $\Lambda$ levels. To capture this effect, we use a phenomenological model of the RPA polarizability $\Pi_\nu(\mathbf{q})$ \cite{hunt_direct_2017,papic_topological_2014,zibrov_tunable_2017},
\begin{equation}
    \Pi^{\chi}(\mathbf{q}) =  \frac{4 \log (4)}{2 \pi} \tanh \left(\chi |\bq|^2 \ell_B^2\right) \frac{1}{\hbar \omega_c}
\end{equation}
where the phenomenological RPA parameter $\chi$ controls the strength of the screening. If we only consider virtual transitions between LLs, matching the small $\bq$ and large $\bq$ behavior of $\Pi_{\nu = 0}(\mathbf{q})$ in Eqn.~\ref{eq:Pi} gives $\chi = 0.62$. Including transitions between $\Lambda$ levels will result in a larger $\chi$, which is turned into a fitting parameter for quasiparticle WC. We note that in DMRG this treatment is unnecessary since it keeps all microscopic electron degrees of freedom, while here we are working directly with quasiparticles.
Using the effective interaction which include both the gate and RPA-screening, the WC energy of Eq.\eqref{eq:WCenergy} is then evaluated numerically as described further in \cite{yang_experimental_2021}. We note that the chemical potential of the quasiparticle $\mu^{\mathrm{qp}}$ computed here is related to the electron chemical potential $\mu$ measured in the experiment by $\mu = \frac{e}{e^*} \mu^{\mathrm{qp}}$.

\subsection{Slow-varying disorder model} 

One tractable limit to account for disorder is when the disorder potential varies slowly in comparison with both the inter-quasiparticle distance in the Wigner crystal and the gate-sample distance. In this limit we can take the local density approximation (LDA) such that the total energy can be expressed as a local functional of the slowly-varying charge density $n(\mathbf{r})$,
\begin{equation}
    E[n] = \int \mathrm{d}^2 \mathcal{E}(n(\mathbf{r})) = \int \mathrm{d}^2  \mathbf{r}\left[\mathcal{E}_{\mathrm{int}}(n(\mathbf{r}))+\frac{e^2 n(\mathbf{r})^2}{2 c_g}+n(\mathbf{r}) e V_D(\mathbf{r})- n(\mathbf{r}) e V_g\right] 
\end{equation}
where $\mathcal{E}_{\mathrm{int}} \equiv \frac{E_{\mathrm{int}}}{2 \pi \ell_B^2 N_{\phi}}$ is the interaction energy density, $c_g = C_g/A$ is the geometric self capacitance per unit area, $V_D(r)$ is the disorder potential, and $V_g$ is the bias on the gate. We note that the disorder model does not care about the microscopic degrees of freedom in the system, which is fully characterized by $\mathcal{E}_{\mathrm{int}}$, so we pick the convention such that $n(\mathbf{r})$ is the electron number density. The key approximation here is the LDA which neglects the gradient $\boldsymbol{\nabla} n$ dependence of the energy functional $E[n]$. Optimizing the energy functional $E_{\mathrm{int}}[n]$ determines the local charge density $n(V_g - V_D)$ at each point,
\begin{equation}
    \mu_{\mathrm{int}}(n(V_g - V_D))+\frac{e^2 n(V_g - V_D)}{c_g} = e V_g - e V_D
\end{equation}
If we consider a disorder distribution $P(V_D)$ with zero mean $\langle V_D \rangle_P = 0$, we find
\begin{equation} \label{eq:nbar}
    \bar{\mu}_{\mathrm{int}}(V_g) + \frac{e^2 \bar{n}(V_g)}{c_g} = e V_g
\end{equation}
where $\bar{\mu}_{\mathrm{int}}(V_g) = \int \mathrm{d} V_D P(V_D) \mu_{\mathrm{int}}(n(V_D))$ is the disorder averaged chemical potential and $\bar{n}(V_g) = \int \mathrm{d} V_D P(V_D) n(V_g - V_D)$ is the average charge density. In the following we will show $\bar{\mu}_{\mathrm{int}}(V_g)$ is precisely the chemical potential measured in the experiment.

As before, the experimentally relevant energy $\bar{E}_{\mathrm{int}}$ is defined relative to a classical capacitor
\begin{equation}
    \bar{E}_{\mathrm{int}}(N_e) = E[n] - \frac{e^2 N_e^2}{2 C_g}+N_e e V_g
\end{equation}
Here the number of electrons $N_e = \bar{n} A$ is determined by the average charge density $\bar{n}$. Then the chemical potential measured in the experiment is
\begin{equation}
    \mu_{\mathrm{expt}}(N_e) = \pdv{\bar{E}_{\mathrm{int}}}{N_e} = \frac1A \int \mathrm{d}^2 \mathbf{r} \pdv{n}{\bar{n}} \pdv{n} \mathcal{E}(n(\mathbf{r})) - \frac{e^2 N_e}{C_g} + e V_g = - \frac{e^2 \bar{n}}{c_g} + e V_g = \bar{\mu}_{\mathrm{int}}(V_g)
\end{equation}
where the integral vanishes because $\partial_n \mathcal{E}(n) = 0$. In practice, we first numerically find the gate bias $V_g$ required to reproduce the electron filling in the experiment using Eq.~\eqref{eq:nbar}, and then compute the disorder averaged chemical potential $\bar{\mu}_{\mathrm{int}}$ assuming a Gaussian disorder distribution $P(V_D)=\frac{1}{\sqrt{2 \pi} \Gamma} \exp (-V_D^2/2\Gamma^2)$.

\begin{table}[h!]
    \renewcommand*{\arraystretch}{1.3}
    \newcolumntype{C}{>{\centering\arraybackslash}X}
    \centering
\begin{tabularx}{0.55\textwidth}{C|CCC} \hline \hline
    & $\Delta$ (\SI{}{meV})                        & $\Gamma$ (\SI{}{meV})                  & $\chi$    \\ \hline
$\displaystyle \nu = -1+1/2$ & $2.4 \pm 0.1$ & $1.0 \pm 0.5$ & $0.85$ \\
$\displaystyle \nu = -1+2/3$ & $3.0 \pm 0.1$ & $1.2 \pm 0.3$ & $0.92$ \\
\hline \hline
\end{tabularx}
\caption{Fitting parameters of the disordered Wigner crystal model at $1/2$ and $2/3$ filling in the $N=1$ LL.}
    \label{tab:parameter} 
\end{table}

The chemical potential $\mu_{\mathrm{int}}$ we have discussed so far sorely comes from the quasiparticle Wigner crystal, where at the FQH fillings $\nu_0$ there is an additional contribution from the correlated FQH gaps,
\begin{equation}
    \mu_{\mathrm{int}} \to \mu_{\mathrm{int}}^{\mathrm{FQH}} = \begin{cases}
        \frac12 \Delta + \mu_{\mathrm{int}}(\nu - \nu_0), \quad &\nu > \nu_0\\
        -\frac12\Delta + \mu_{\mathrm{int}}(\nu - \nu_0), \quad &\nu < \nu_0
    \end{cases}
\end{equation}
where $\Delta = \frac{e}{e^*} \Delta_{\mathrm{qp}}$ is the thermodynamic gap at $\nu_0$. The quasiparticle Wigner crystal model does not break the particle hole symmetry, so we can obtain the chemical potential at $\nu < \nu_0$ by $\mu_{\mathrm{int}}(\Delta \nu) = - \mu_{\mathrm{int}}(-\Delta \nu)$. As shown in Fig.~\ref{fig:2}, we fit the disorder-averaged chemical potential $\bar{\mu}_{\mathrm{int}}^{\mathrm{FQH}}$ to the experiment data with three free parameters: the thermodynamic gap $\Delta$, the disorder scale $\Gamma$ and the screening parameter $\chi$. We summarize these fitting parameters at $\nu = -1 + 1/2$ and $-1 + 2/3$ in the N=1 LL in Table~\ref{tab:parameter}.

\section{Extended data and figures}

\begin{figure*}[ht]
    \centering
    \includegraphics[width = 90mm]{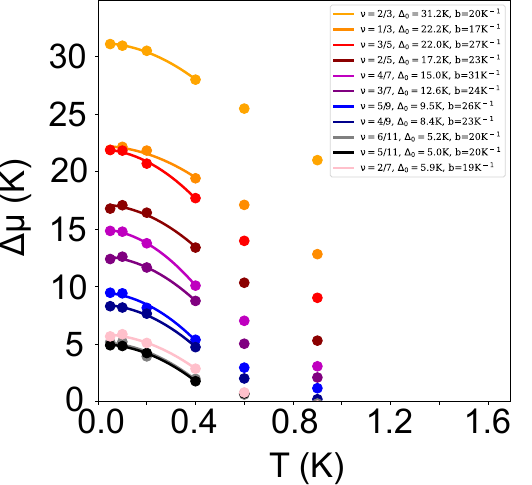}
    \caption{Chemical potential jump as a function of temperature for the Jain states in N=0 at $B=\SI{13.8}{T}$. The data (dots) are fitted using the low temperature Sommerfeld expansion (solid lines).}
    \label{fig:SomJain}
\end{figure*}

\begin{table}[h]
    \renewcommand*{\arraystretch}{1.3}
    \newcolumntype{C}{>{\centering\arraybackslash}X}
    \centering
\begin{tabularx}{0.48\textwidth}{|C|C|C||C|}
\hline
Filling $\nu$ & $\Delta_{\mathrm{qp}}^{\mathrm{\mu}}$ (K) & $\Delta_{\mathrm{qp}}^{\mathrm{fit}}$ (K) & $\Delta_{\mathrm{qp}}^{\mathrm{DMRG}}$ (K) \\ \hline
$1/3$ & \SI{7.6}{K} & \SI{17.7}{K} & \SI{13.5}{K} \\ \hline
$2/5$ & \SI{3.6}{K} & \SI{7.7}{K} & \SI{6.3}{K} \\ \hline
$3/7$ & \SI{1.9}{K} & \SI{5.7}{K} & \SI{4.5}{K} \\ \hline
$4/9$ & \SI{0.9}{K} & \SI{3.4}{K} & $-$ \\ \hline
$5/11$ & \SI{0.4}{K} & \SI{2.1}{K} & $-$ \\ \hline
$6/11$ & \SI{0.4}{K} & \SI{2.1}{K} & $-$ \\ \hline
$5/9$ & \SI{1.0}{K} & \SI{3.5}{K} & $-$ \\ \hline
$4/7$ & \SI{2.1}{K} & \SI{6.0}{K} & $-$ \\ \hline
$3/5$ & \SI{4.4}{K} & \SI{10.5}{K} & $-$ \\ \hline
$2/3$ & \SI{10.5}{K} & \SI{21.2}{K} & \SI{14.3}{K} \\ \hline
$2/7$ & \SI{0.9}{K} & \SI{4.3}{K} & $-$ \\ \hline
    \end{tabularx}
    \caption{Comparison of the quasiparticle gaps at the Jain sequence in the $N = 0$ LL at $B = \SI{13.8}{T}$ as determined by DMRG calculations $\Delta_{\mathrm{qp}}^{\mathrm{DMRG}}$, the chemical potential jump $\Delta_{\mathrm{qp}}^{\mu}$, and from the fit to the Wigner crystal model $\Delta_{\mathrm{qp}}^{\mathrm{fit}}$. }
    \label{tab:Jain}
\end{table}


\begin{thebibliography}{47}%
\makeatletter
\providecommand \@ifxundefined [1]{%
 \@ifx{#1\undefined}
}%
\providecommand \@ifnum [1]{%
 \ifnum #1\expandafter \@firstoftwo
 \else \expandafter \@secondoftwo
 \fi
}%
\providecommand \@ifx [1]{%
 \ifx #1\expandafter \@firstoftwo
 \else \expandafter \@secondoftwo
 \fi
}%
\providecommand \natexlab [1]{#1}%
\providecommand \enquote  [1]{``#1''}%
\providecommand \bibnamefont  [1]{#1}%
\providecommand \bibfnamefont [1]{#1}%
\providecommand \citenamefont [1]{#1}%
\providecommand \href@noop [0]{\@secondoftwo}%
\providecommand \href [0]{\begingroup \@sanitize@url \@href}%
\providecommand \@href[1]{\@@startlink{#1}\@@href}%
\providecommand \@@href[1]{\endgroup#1\@@endlink}%
\providecommand \@sanitize@url [0]{\catcode `\\12\catcode `\$12\catcode
  `\&12\catcode `\#12\catcode `\^12\catcode `\_12\catcode `\%12\relax}%
\providecommand \@@startlink[1]{}%
\providecommand \@@endlink[0]{}%
\providecommand \url  [0]{\begingroup\@sanitize@url \@url }%
\providecommand \@url [1]{\endgroup\@href {#1}{\urlprefix }}%
\providecommand \urlprefix  [0]{URL }%
\providecommand \Eprint [0]{\href }%
\providecommand \doibase [0]{http://dx.doi.org/}%
\providecommand \selectlanguage [0]{\@gobble}%
\providecommand \bibinfo  [0]{\@secondoftwo}%
\providecommand \bibfield  [0]{\@secondoftwo}%
\providecommand \translation [1]{[#1]}%
\providecommand \BibitemOpen [0]{}%
\providecommand \bibitemStop [0]{}%
\providecommand \bibitemNoStop [0]{.\EOS\space}%
\providecommand \EOS [0]{\spacefactor3000\relax}%
\providecommand \BibitemShut  [1]{\csname bibitem#1\endcsname}%
\let\auto@bib@innerbib\@empty
\bibitem [{\citenamefont {Moore}\ and\ \citenamefont
  {Read}(1991)}]{moore_nonabelions_1991}%
  \BibitemOpen
  \bibfield  {author} {\bibinfo {author} {\bibfnamefont {G.}~\bibnamefont
  {Moore}}\ and\ \bibinfo {author} {\bibfnamefont {N.}~\bibnamefont {Read}},\
  }\href {http://www.sciencedirect.com/science/article/pii/055032139190407O}
  {\bibfield  {journal} {\bibinfo  {journal} {Nuclear Physics B}\ }\textbf
  {\bibinfo {volume} {360}},\ \bibinfo {pages} {362} (\bibinfo {year}
  {1991})}\BibitemShut {NoStop}%
\bibitem [{\citenamefont {Nayak}\ \emph {et~al.}(2008)\citenamefont {Nayak},
  \citenamefont {Simon}, \citenamefont {Stern}, \citenamefont {Freedman},\ and\
  \citenamefont {Das~Sarma}}]{nayak_non-abelian_2008}%
  \BibitemOpen
  \bibfield  {author} {\bibinfo {author} {\bibfnamefont {C.}~\bibnamefont
  {Nayak}}, \bibinfo {author} {\bibfnamefont {S.~H.}\ \bibnamefont {Simon}},
  \bibinfo {author} {\bibfnamefont {A.}~\bibnamefont {Stern}}, \bibinfo
  {author} {\bibfnamefont {M.}~\bibnamefont {Freedman}}, \ and\ \bibinfo
  {author} {\bibfnamefont {S.}~\bibnamefont {Das~Sarma}},\ }\href
  {http://link.aps.org/doi/10.1103/RevModPhys.80.1083} {\bibfield  {journal}
  {\bibinfo  {journal} {Reviews of Modern Physics}\ }\textbf {\bibinfo {volume}
  {80}},\ \bibinfo {pages} {1083} (\bibinfo {year} {2008})}\BibitemShut
  {NoStop}%
\bibitem [{\citenamefont {Morf}(1998)}]{morf_transition_1998}%
  \BibitemOpen
  \bibfield  {author} {\bibinfo {author} {\bibfnamefont {R.~H.}\ \bibnamefont
  {Morf}},\ }\href {\doibase 10.1103/PhysRevLett.80.1505} {\bibfield  {journal}
  {\bibinfo  {journal} {Physical Review Letters}\ }\textbf {\bibinfo {volume}
  {80}},\ \bibinfo {pages} {1505} (\bibinfo {year} {1998})}\BibitemShut
  {NoStop}%
\bibitem [{\citenamefont {Rezayi}(2017)}]{rezayi_landau-level-mixing_2017}%
  \BibitemOpen
  \bibfield  {author} {\bibinfo {author} {\bibfnamefont {E.~H.}\ \bibnamefont
  {Rezayi}},\ }\href {http://arxiv.org/abs/1704.03026} {\bibfield  {journal}
  {\bibinfo  {journal} {arXiv:1704.03026 [cond-mat]}\ } (\bibinfo {year}
  {2017})},\ \bibinfo {note} {arXiv: 1704.03026}\BibitemShut {NoStop}%
\bibitem [{\citenamefont {Willett}\ \emph {et~al.}(1987)\citenamefont
  {Willett}, \citenamefont {Eisenstein}, \citenamefont {Stormer}, \citenamefont
  {Tsui}, \citenamefont {Gossard},\ and\ \citenamefont
  {English}}]{willett_observation_1987}%
  \BibitemOpen
  \bibfield  {author} {\bibinfo {author} {\bibfnamefont {R.}~\bibnamefont
  {Willett}}, \bibinfo {author} {\bibfnamefont {J.~P.}\ \bibnamefont
  {Eisenstein}}, \bibinfo {author} {\bibfnamefont {H.~L.}\ \bibnamefont
  {Stormer}}, \bibinfo {author} {\bibfnamefont {D.~C.}\ \bibnamefont {Tsui}},
  \bibinfo {author} {\bibfnamefont {A.~C.}\ \bibnamefont {Gossard}}, \ and\
  \bibinfo {author} {\bibfnamefont {J.~H.}\ \bibnamefont {English}},\ }\href
  {http://link.aps.org/doi/10.1103/PhysRevLett.59.1776} {\bibfield  {journal}
  {\bibinfo  {journal} {Phys. Rev. Lett.}\ }\textbf {\bibinfo {volume} {59}}
  (\bibinfo {year} {1987})}\BibitemShut {NoStop}%
\bibitem [{\citenamefont {Kumar}\ \emph {et~al.}(2010)\citenamefont {Kumar},
  \citenamefont {Csáthy}, \citenamefont {Manfra}, \citenamefont {Pfeiffer},\
  and\ \citenamefont {West}}]{kumar_nonconventional_2010}%
  \BibitemOpen
  \bibfield  {author} {\bibinfo {author} {\bibfnamefont {A.}~\bibnamefont
  {Kumar}}, \bibinfo {author} {\bibfnamefont {G.~A.}\ \bibnamefont {Csáthy}},
  \bibinfo {author} {\bibfnamefont {M.~J.}\ \bibnamefont {Manfra}}, \bibinfo
  {author} {\bibfnamefont {L.~N.}\ \bibnamefont {Pfeiffer}}, \ and\ \bibinfo
  {author} {\bibfnamefont {K.~W.}\ \bibnamefont {West}},\ }\href {\doibase
  10.1103/PhysRevLett.105.246808} {\bibfield  {journal} {\bibinfo  {journal}
  {Physical Review Letters}\ }\textbf {\bibinfo {volume} {105}},\ \bibinfo
  {pages} {246808} (\bibinfo {year} {2010})}\BibitemShut {NoStop}%
\bibitem [{\citenamefont {Watson}\ \emph {et~al.}(2015)\citenamefont {Watson},
  \citenamefont {Csáthy},\ and\ \citenamefont {Manfra}}]{watson_impact_2015}%
  \BibitemOpen
  \bibfield  {author} {\bibinfo {author} {\bibfnamefont {J.}~\bibnamefont
  {Watson}}, \bibinfo {author} {\bibfnamefont {G.}~\bibnamefont {Csáthy}}, \
  and\ \bibinfo {author} {\bibfnamefont {M.}~\bibnamefont {Manfra}},\ }\href
  {\doibase 10.1103/PhysRevApplied.3.064004} {\bibfield  {journal} {\bibinfo
  {journal} {Physical Review Applied}\ }\textbf {\bibinfo {volume} {3}},\
  \bibinfo {pages} {064004} (\bibinfo {year} {2015})},\ \bibinfo {note}
  {publisher: American Physical Society}\BibitemShut {NoStop}%
\bibitem [{\citenamefont {Chung}\ \emph {et~al.}(2021)\citenamefont {Chung},
  \citenamefont {Villegas~Rosales}, \citenamefont {Baldwin}, \citenamefont
  {Madathil}, \citenamefont {West}, \citenamefont {Shayegan},\ and\
  \citenamefont {Pfeiffer}}]{chung_ultra-high-quality_2021}%
  \BibitemOpen
  \bibfield  {author} {\bibinfo {author} {\bibfnamefont {Y.~J.}\ \bibnamefont
  {Chung}}, \bibinfo {author} {\bibfnamefont {K.~A.}\ \bibnamefont
  {Villegas~Rosales}}, \bibinfo {author} {\bibfnamefont {K.~W.}\ \bibnamefont
  {Baldwin}}, \bibinfo {author} {\bibfnamefont {P.~T.}\ \bibnamefont
  {Madathil}}, \bibinfo {author} {\bibfnamefont {K.~W.}\ \bibnamefont {West}},
  \bibinfo {author} {\bibfnamefont {M.}~\bibnamefont {Shayegan}}, \ and\
  \bibinfo {author} {\bibfnamefont {L.~N.}\ \bibnamefont {Pfeiffer}},\ }\href
  {\doibase 10.1038/s41563-021-00942-3} {\bibfield  {journal} {\bibinfo
  {journal} {Nature Materials}\ }\textbf {\bibinfo {volume} {20}},\ \bibinfo
  {pages} {632} (\bibinfo {year} {2021})},\ \bibinfo {note} {number: 5
  Publisher: Nature Publishing Group}\BibitemShut {NoStop}%
\bibitem [{\citenamefont {McCann}\ and\ \citenamefont
  {Fal'ko}(2006)}]{mccann_landau-level_2006}%
  \BibitemOpen
  \bibfield  {author} {\bibinfo {author} {\bibfnamefont {E.}~\bibnamefont
  {McCann}}\ and\ \bibinfo {author} {\bibfnamefont {V.~I.}\ \bibnamefont
  {Fal'ko}},\ }\href {\doibase 10.1103/PhysRevLett.96.086805} {\bibfield
  {journal} {\bibinfo  {journal} {Phys. Rev. Lett.}\ }\textbf {\bibinfo
  {volume} {96}} (\bibinfo {year} {2006}),\
  10.1103/PhysRevLett.96.086805}\BibitemShut {NoStop}%
\bibitem [{\citenamefont {Dean}\ \emph {et~al.}(2020)\citenamefont {Dean},
  \citenamefont {Kim}, \citenamefont {Li},\ and\ \citenamefont
  {Young}}]{dean_fractional_2020}%
  \BibitemOpen
  \bibfield  {author} {\bibinfo {author} {\bibfnamefont {C.}~\bibnamefont
  {Dean}}, \bibinfo {author} {\bibfnamefont {P.}~\bibnamefont {Kim}}, \bibinfo
  {author} {\bibfnamefont {J.~I.~A.}\ \bibnamefont {Li}}, \ and\ \bibinfo
  {author} {\bibfnamefont {A.}~\bibnamefont {Young}},\ }in\ \href {\doibase
  10.1142/9789811217494_0007} {\emph {\bibinfo {booktitle} {Fractional
  {Quantum} {Hall} {Effects}: {New} {Developments}}}}\ (\bibinfo  {publisher}
  {World Scientific},\ \bibinfo {address} {Singapore},\ \bibinfo {year}
  {2020})\ pp.\ \bibinfo {pages} {317--375}\BibitemShut {NoStop}%
\bibitem [{\citenamefont {Feldman}\ \emph {et~al.}(2009)\citenamefont
  {Feldman}, \citenamefont {Martin},\ and\ \citenamefont
  {Yacoby}}]{feldman_broken-symmetry_2009}%
  \BibitemOpen
  \bibfield  {author} {\bibinfo {author} {\bibfnamefont {B.~E.}\ \bibnamefont
  {Feldman}}, \bibinfo {author} {\bibfnamefont {J.}~\bibnamefont {Martin}}, \
  and\ \bibinfo {author} {\bibfnamefont {A.}~\bibnamefont {Yacoby}},\ }\href
  {http://dx.doi.org/10.1038/nphys1406} {\bibfield  {journal} {\bibinfo
  {journal} {Nature Physics}\ }\textbf {\bibinfo {volume} {5}},\ \bibinfo
  {pages} {889} (\bibinfo {year} {2009})}\BibitemShut {NoStop}%
\bibitem [{\citenamefont {Martin}\ \emph {et~al.}(2010)\citenamefont {Martin},
  \citenamefont {Feldman}, \citenamefont {Weitz}, \citenamefont {Allen},\ and\
  \citenamefont {Yacoby}}]{martin_local_2010}%
  \BibitemOpen
  \bibfield  {author} {\bibinfo {author} {\bibfnamefont {J.}~\bibnamefont
  {Martin}}, \bibinfo {author} {\bibfnamefont {B.~E.}\ \bibnamefont {Feldman}},
  \bibinfo {author} {\bibfnamefont {R.~T.}\ \bibnamefont {Weitz}}, \bibinfo
  {author} {\bibfnamefont {M.~T.}\ \bibnamefont {Allen}}, \ and\ \bibinfo
  {author} {\bibfnamefont {A.}~\bibnamefont {Yacoby}},\ }\href
  {http://link.aps.org/doi/10.1103/PhysRevLett.105.256806} {\bibfield
  {journal} {\bibinfo  {journal} {Phys. Rev. Lett.}\ }\textbf {\bibinfo
  {volume} {105}} (\bibinfo {year} {2010})}\BibitemShut {NoStop}%
\bibitem [{\citenamefont {Dean}\ \emph {et~al.}(2010)\citenamefont {Dean},
  \citenamefont {Young}, \citenamefont {Meric}, \citenamefont {Lee},
  \citenamefont {Wang}, \citenamefont {Sorgenfrei}, \citenamefont {Watanabe},
  \citenamefont {Taniguchi}, \citenamefont {Kim}, \citenamefont {Shepard},\
  and\ \citenamefont {Hone}}]{dean_boron_2010}%
  \BibitemOpen
  \bibfield  {author} {\bibinfo {author} {\bibfnamefont {C.~R.}\ \bibnamefont
  {Dean}}, \bibinfo {author} {\bibfnamefont {A.~F.}\ \bibnamefont {Young}},
  \bibinfo {author} {\bibfnamefont {I.}~\bibnamefont {Meric}}, \bibinfo
  {author} {\bibfnamefont {C.}~\bibnamefont {Lee}}, \bibinfo {author}
  {\bibfnamefont {L.}~\bibnamefont {Wang}}, \bibinfo {author} {\bibfnamefont
  {S.}~\bibnamefont {Sorgenfrei}}, \bibinfo {author} {\bibfnamefont
  {K.}~\bibnamefont {Watanabe}}, \bibinfo {author} {\bibfnamefont
  {T.}~\bibnamefont {Taniguchi}}, \bibinfo {author} {\bibfnamefont
  {P.}~\bibnamefont {Kim}}, \bibinfo {author} {\bibfnamefont {K.~L.}\
  \bibnamefont {Shepard}}, \ and\ \bibinfo {author} {\bibfnamefont
  {J.}~\bibnamefont {Hone}},\ }\href {http://dx.doi.org/10.1038/nnano.2010.172}
  {\bibfield  {journal} {\bibinfo  {journal} {Nature Nanotechnology}\ }\textbf
  {\bibinfo {volume} {5}},\ \bibinfo {pages} {722} (\bibinfo {year}
  {2010})}\BibitemShut {NoStop}%
\bibitem [{\citenamefont {Lee}\ \emph {et~al.}(2014)\citenamefont {Lee},
  \citenamefont {Fallahazad}, \citenamefont {Xue}, \citenamefont {Dillen},
  \citenamefont {Kim}, \citenamefont {Taniguchi}, \citenamefont {Watanabe},\
  and\ \citenamefont {Tutuc}}]{lee_chemical_2014}%
  \BibitemOpen
  \bibfield  {author} {\bibinfo {author} {\bibfnamefont {K.}~\bibnamefont
  {Lee}}, \bibinfo {author} {\bibfnamefont {B.}~\bibnamefont {Fallahazad}},
  \bibinfo {author} {\bibfnamefont {J.}~\bibnamefont {Xue}}, \bibinfo {author}
  {\bibfnamefont {D.~C.}\ \bibnamefont {Dillen}}, \bibinfo {author}
  {\bibfnamefont {K.}~\bibnamefont {Kim}}, \bibinfo {author} {\bibfnamefont
  {T.}~\bibnamefont {Taniguchi}}, \bibinfo {author} {\bibfnamefont
  {K.}~\bibnamefont {Watanabe}}, \ and\ \bibinfo {author} {\bibfnamefont
  {E.}~\bibnamefont {Tutuc}},\ }\href {\doibase 10.1126/science.1251003}
  {\bibfield  {journal} {\bibinfo  {journal} {Science}\ }\textbf {\bibinfo
  {volume} {345}},\ \bibinfo {pages} {58} (\bibinfo {year} {2014})}\BibitemShut
  {NoStop}%
\bibitem [{\citenamefont {Kou}\ \emph {et~al.}(2014)\citenamefont {Kou},
  \citenamefont {Feldman}, \citenamefont {Levin}, \citenamefont {Halperin},
  \citenamefont {Watanabe}, \citenamefont {Taniguchi},\ and\ \citenamefont
  {Yacoby}}]{kou_electron-hole_2014}%
  \BibitemOpen
  \bibfield  {author} {\bibinfo {author} {\bibfnamefont {A.}~\bibnamefont
  {Kou}}, \bibinfo {author} {\bibfnamefont {B.~E.}\ \bibnamefont {Feldman}},
  \bibinfo {author} {\bibfnamefont {A.~J.}\ \bibnamefont {Levin}}, \bibinfo
  {author} {\bibfnamefont {B.~I.}\ \bibnamefont {Halperin}}, \bibinfo {author}
  {\bibfnamefont {K.}~\bibnamefont {Watanabe}}, \bibinfo {author}
  {\bibfnamefont {T.}~\bibnamefont {Taniguchi}}, \ and\ \bibinfo {author}
  {\bibfnamefont {A.}~\bibnamefont {Yacoby}},\ }\href {\doibase
  10.1126/science.1250270} {\bibfield  {journal} {\bibinfo  {journal}
  {Science}\ }\textbf {\bibinfo {volume} {345}},\ \bibinfo {pages} {55}
  (\bibinfo {year} {2014})}\BibitemShut {NoStop}%
\bibitem [{\citenamefont {Ki}\ \emph {et~al.}(2014)\citenamefont {Ki},
  \citenamefont {Fal’ko}, \citenamefont {Abanin},\ and\ \citenamefont
  {Morpurgo}}]{ki_observation_2014}%
  \BibitemOpen
  \bibfield  {author} {\bibinfo {author} {\bibfnamefont {D.-K.}\ \bibnamefont
  {Ki}}, \bibinfo {author} {\bibfnamefont {V.~I.}\ \bibnamefont {Fal’ko}},
  \bibinfo {author} {\bibfnamefont {D.~A.}\ \bibnamefont {Abanin}}, \ and\
  \bibinfo {author} {\bibfnamefont {A.~F.}\ \bibnamefont {Morpurgo}},\ }\href
  {\doibase 10.1021/nl5003922} {\bibfield  {journal} {\bibinfo  {journal} {Nano
  Letters}\ }\textbf {\bibinfo {volume} {14}},\ \bibinfo {pages} {2135}
  (\bibinfo {year} {2014})}\BibitemShut {NoStop}%
\bibitem [{\citenamefont {Maher}\ \emph {et~al.}(2014)\citenamefont {Maher},
  \citenamefont {Wang}, \citenamefont {Gao}, \citenamefont {Forsythe},
  \citenamefont {Taniguchi}, \citenamefont {Watanabe}, \citenamefont {Abanin},
  \citenamefont {Papic}, \citenamefont {Cadden-Zimansky}, \citenamefont {Hone},
  \citenamefont {Kim},\ and\ \citenamefont {Dean}}]{maher_tunable_2014}%
  \BibitemOpen
  \bibfield  {author} {\bibinfo {author} {\bibfnamefont {P.}~\bibnamefont
  {Maher}}, \bibinfo {author} {\bibfnamefont {L.}~\bibnamefont {Wang}},
  \bibinfo {author} {\bibfnamefont {Y.}~\bibnamefont {Gao}}, \bibinfo {author}
  {\bibfnamefont {C.}~\bibnamefont {Forsythe}}, \bibinfo {author}
  {\bibfnamefont {T.}~\bibnamefont {Taniguchi}}, \bibinfo {author}
  {\bibfnamefont {K.}~\bibnamefont {Watanabe}}, \bibinfo {author}
  {\bibfnamefont {D.}~\bibnamefont {Abanin}}, \bibinfo {author} {\bibfnamefont
  {Z.}~\bibnamefont {Papic}}, \bibinfo {author} {\bibfnamefont
  {P.}~\bibnamefont {Cadden-Zimansky}}, \bibinfo {author} {\bibfnamefont
  {J.}~\bibnamefont {Hone}}, \bibinfo {author} {\bibfnamefont {P.}~\bibnamefont
  {Kim}}, \ and\ \bibinfo {author} {\bibfnamefont {C.~R.}\ \bibnamefont
  {Dean}},\ }\href {\doibase 10.1126/science.1252875} {\bibfield  {journal}
  {\bibinfo  {journal} {Science}\ }\textbf {\bibinfo {volume} {345}},\ \bibinfo
  {pages} {61} (\bibinfo {year} {2014})}\BibitemShut {NoStop}%
\bibitem [{\citenamefont {Hunt}\ \emph {et~al.}(2017)\citenamefont {Hunt},
  \citenamefont {Li}, \citenamefont {Zibrov}, \citenamefont {Wang},
  \citenamefont {Taniguchi}, \citenamefont {Watanabe}, \citenamefont {Hone},
  \citenamefont {Dean}, \citenamefont {Zaletel}, \citenamefont {Ashoori},\ and\
  \citenamefont {Young}}]{hunt_direct_2017}%
  \BibitemOpen
  \bibfield  {author} {\bibinfo {author} {\bibfnamefont {B.~M.}\ \bibnamefont
  {Hunt}}, \bibinfo {author} {\bibfnamefont {J.~I.~A.}\ \bibnamefont {Li}},
  \bibinfo {author} {\bibfnamefont {A.~A.}\ \bibnamefont {Zibrov}}, \bibinfo
  {author} {\bibfnamefont {L.}~\bibnamefont {Wang}}, \bibinfo {author}
  {\bibfnamefont {T.}~\bibnamefont {Taniguchi}}, \bibinfo {author}
  {\bibfnamefont {K.}~\bibnamefont {Watanabe}}, \bibinfo {author}
  {\bibfnamefont {J.}~\bibnamefont {Hone}}, \bibinfo {author} {\bibfnamefont
  {C.~R.}\ \bibnamefont {Dean}}, \bibinfo {author} {\bibfnamefont
  {M.}~\bibnamefont {Zaletel}}, \bibinfo {author} {\bibfnamefont {R.~C.}\
  \bibnamefont {Ashoori}}, \ and\ \bibinfo {author} {\bibfnamefont {A.~F.}\
  \bibnamefont {Young}},\ }\href {\doibase 10.1038/s41467-017-00824-w}
  {\bibfield  {journal} {\bibinfo  {journal} {Nature Communications}\ }\textbf
  {\bibinfo {volume} {8}},\ \bibinfo {pages} {948} (\bibinfo {year}
  {2017})}\BibitemShut {NoStop}%
\bibitem [{\citenamefont {Huang}\ \emph {et~al.}(2022)\citenamefont {Huang},
  \citenamefont {Fu}, \citenamefont {Hickey}, \citenamefont {Alem},
  \citenamefont {Lin}, \citenamefont {Watanabe}, \citenamefont {Taniguchi},\
  and\ \citenamefont {Zhu}}]{huang_valley_2022}%
  \BibitemOpen
  \bibfield  {author} {\bibinfo {author} {\bibfnamefont {K.}~\bibnamefont
  {Huang}}, \bibinfo {author} {\bibfnamefont {H.}~\bibnamefont {Fu}}, \bibinfo
  {author} {\bibfnamefont {D.~R.}\ \bibnamefont {Hickey}}, \bibinfo {author}
  {\bibfnamefont {N.}~\bibnamefont {Alem}}, \bibinfo {author} {\bibfnamefont
  {X.}~\bibnamefont {Lin}}, \bibinfo {author} {\bibfnamefont {K.}~\bibnamefont
  {Watanabe}}, \bibinfo {author} {\bibfnamefont {T.}~\bibnamefont {Taniguchi}},
  \ and\ \bibinfo {author} {\bibfnamefont {J.}~\bibnamefont {Zhu}},\ }\href
  {\doibase 10.1103/PhysRevX.12.031019} {\bibfield  {journal} {\bibinfo
  {journal} {Physical Review X}\ }\textbf {\bibinfo {volume} {12}},\ \bibinfo
  {pages} {031019} (\bibinfo {year} {2022})},\ \bibinfo {note} {publisher:
  American Physical Society}\BibitemShut {NoStop}%
\bibitem [{\citenamefont {Zibrov}\ \emph {et~al.}(2017)\citenamefont {Zibrov},
  \citenamefont {Kometter}, \citenamefont {Zhou}, \citenamefont {Spanton},
  \citenamefont {Taniguchi}, \citenamefont {Watanabe}, \citenamefont
  {Zaletel},\ and\ \citenamefont {Young}}]{zibrov_tunable_2017}%
  \BibitemOpen
  \bibfield  {author} {\bibinfo {author} {\bibfnamefont {A.~A.}\ \bibnamefont
  {Zibrov}}, \bibinfo {author} {\bibfnamefont {C.}~\bibnamefont {Kometter}},
  \bibinfo {author} {\bibfnamefont {H.}~\bibnamefont {Zhou}}, \bibinfo {author}
  {\bibfnamefont {E.~M.}\ \bibnamefont {Spanton}}, \bibinfo {author}
  {\bibfnamefont {T.}~\bibnamefont {Taniguchi}}, \bibinfo {author}
  {\bibfnamefont {K.}~\bibnamefont {Watanabe}}, \bibinfo {author}
  {\bibfnamefont {M.~P.}\ \bibnamefont {Zaletel}}, \ and\ \bibinfo {author}
  {\bibfnamefont {A.~F.}\ \bibnamefont {Young}},\ }\href {\doibase
  10.1038/nature23893} {\bibfield  {journal} {\bibinfo  {journal} {Nature}\
  }\textbf {\bibinfo {volume} {549}},\ \bibinfo {pages} {360} (\bibinfo {year}
  {2017})}\BibitemShut {NoStop}%
\bibitem [{\citenamefont {Li}\ \emph {et~al.}(2017)\citenamefont {Li},
  \citenamefont {Tan}, \citenamefont {Chen}, \citenamefont {Zeng},
  \citenamefont {Taniguchi}, \citenamefont {Watanabe}, \citenamefont {Hone},\
  and\ \citenamefont {Dean}}]{li_even_2017}%
  \BibitemOpen
  \bibfield  {author} {\bibinfo {author} {\bibfnamefont {J.~I.~A.}\
  \bibnamefont {Li}}, \bibinfo {author} {\bibfnamefont {C.}~\bibnamefont
  {Tan}}, \bibinfo {author} {\bibfnamefont {S.}~\bibnamefont {Chen}}, \bibinfo
  {author} {\bibfnamefont {Y.}~\bibnamefont {Zeng}}, \bibinfo {author}
  {\bibfnamefont {T.}~\bibnamefont {Taniguchi}}, \bibinfo {author}
  {\bibfnamefont {K.}~\bibnamefont {Watanabe}}, \bibinfo {author}
  {\bibfnamefont {J.}~\bibnamefont {Hone}}, \ and\ \bibinfo {author}
  {\bibfnamefont {C.~R.}\ \bibnamefont {Dean}},\ }\href {\doibase
  10.1126/science.aao2521} {\bibfield  {journal} {\bibinfo  {journal}
  {Science}\ ,\ \bibinfo {pages} {eaao2521}} (\bibinfo {year}
  {2017})}\BibitemShut {NoStop}%
\bibitem [{\citenamefont {Apalkov}\ and\ \citenamefont
  {Chakraborty}(2011)}]{apalkov_stable_2011}%
  \BibitemOpen
  \bibfield  {author} {\bibinfo {author} {\bibfnamefont {V.~M.}\ \bibnamefont
  {Apalkov}}\ and\ \bibinfo {author} {\bibfnamefont {T.}~\bibnamefont
  {Chakraborty}},\ }\href {\doibase 10.1103/PhysRevLett.107.186803} {\bibfield
  {journal} {\bibinfo  {journal} {Physical Review Letters}\ }\textbf {\bibinfo
  {volume} {107}},\ \bibinfo {pages} {186803} (\bibinfo {year}
  {2011})}\BibitemShut {NoStop}%
\bibitem [{\citenamefont {Papic}\ and\ \citenamefont
  {Abanin}(2014)}]{papic_topological_2014}%
  \BibitemOpen
  \bibfield  {author} {\bibinfo {author} {\bibfnamefont {Z.}~\bibnamefont
  {Papic}}\ and\ \bibinfo {author} {\bibfnamefont {D.~A.}\ \bibnamefont
  {Abanin}},\ }\href {\doibase 10.1103/PhysRevLett.112.046602} {\bibfield
  {journal} {\bibinfo  {journal} {Physical Review Letters}\ }\textbf {\bibinfo
  {volume} {112}},\ \bibinfo {pages} {046602} (\bibinfo {year}
  {2014})}\BibitemShut {NoStop}%
\bibitem [{\citenamefont {Balram}(2022)}]{balram_transitions_2022}%
  \BibitemOpen
  \bibfield  {author} {\bibinfo {author} {\bibfnamefont {A.~C.}\ \bibnamefont
  {Balram}},\ }\href {\doibase 10.1103/PhysRevB.105.L121406} {\bibfield
  {journal} {\bibinfo  {journal} {Physical Review B}\ }\textbf {\bibinfo
  {volume} {105}},\ \bibinfo {pages} {L121406} (\bibinfo {year} {2022})},\
  \bibinfo {note} {publisher: American Physical Society}\BibitemShut {NoStop}%
\bibitem [{\citenamefont {Polshyn}\ \emph {et~al.}(2018)\citenamefont
  {Polshyn}, \citenamefont {Zhou}, \citenamefont {Spanton}, \citenamefont
  {Taniguchi}, \citenamefont {Watanabe},\ and\ \citenamefont
  {Young}}]{polshyn_quantitative_2018}%
  \BibitemOpen
  \bibfield  {author} {\bibinfo {author} {\bibfnamefont {H.}~\bibnamefont
  {Polshyn}}, \bibinfo {author} {\bibfnamefont {H.}~\bibnamefont {Zhou}},
  \bibinfo {author} {\bibfnamefont {E.~M.}\ \bibnamefont {Spanton}}, \bibinfo
  {author} {\bibfnamefont {T.}~\bibnamefont {Taniguchi}}, \bibinfo {author}
  {\bibfnamefont {K.}~\bibnamefont {Watanabe}}, \ and\ \bibinfo {author}
  {\bibfnamefont {A.~F.}\ \bibnamefont {Young}},\ }\href {\doibase
  10.1103/PhysRevLett.121.226801} {\bibfield  {journal} {\bibinfo  {journal}
  {Physical Review Letters}\ }\textbf {\bibinfo {volume} {121}},\ \bibinfo
  {pages} {226801} (\bibinfo {year} {2018})}\BibitemShut {NoStop}%
\bibitem [{\citenamefont {Zeng}\ \emph {et~al.}(2019)\citenamefont {Zeng},
  \citenamefont {Li}, \citenamefont {Dietrich}, \citenamefont {Ghosh},
  \citenamefont {Watanabe}, \citenamefont {Taniguchi}, \citenamefont {Hone},\
  and\ \citenamefont {Dean}}]{zeng_high-quality_2019}%
  \BibitemOpen
  \bibfield  {author} {\bibinfo {author} {\bibfnamefont {Y.}~\bibnamefont
  {Zeng}}, \bibinfo {author} {\bibfnamefont {J.~I.~A.}\ \bibnamefont {Li}},
  \bibinfo {author} {\bibfnamefont {S.~A.}\ \bibnamefont {Dietrich}}, \bibinfo
  {author} {\bibfnamefont {O.~M.}\ \bibnamefont {Ghosh}}, \bibinfo {author}
  {\bibfnamefont {K.}~\bibnamefont {Watanabe}}, \bibinfo {author}
  {\bibfnamefont {T.}~\bibnamefont {Taniguchi}}, \bibinfo {author}
  {\bibfnamefont {J.}~\bibnamefont {Hone}}, \ and\ \bibinfo {author}
  {\bibfnamefont {C.~R.}\ \bibnamefont {Dean}},\ }\href {\doibase
  10.1103/PhysRevLett.122.137701} {\bibfield  {journal} {\bibinfo  {journal}
  {Physical Review Letters}\ }\textbf {\bibinfo {volume} {122}},\ \bibinfo
  {pages} {137701} (\bibinfo {year} {2019})}\BibitemShut {NoStop}%
\bibitem [{\citenamefont {Eisenstein}\ \emph {et~al.}(1994)\citenamefont
  {Eisenstein}, \citenamefont {Pfeiffer},\ and\ \citenamefont
  {West}}]{eisenstein_compressibility_1994}%
  \BibitemOpen
  \bibfield  {author} {\bibinfo {author} {\bibfnamefont {J.~P.}\ \bibnamefont
  {Eisenstein}}, \bibinfo {author} {\bibfnamefont {L.~N.}\ \bibnamefont
  {Pfeiffer}}, \ and\ \bibinfo {author} {\bibfnamefont {K.~W.}\ \bibnamefont
  {West}},\ }\href {\doibase 10.1103/PhysRevB.50.1760} {\bibfield  {journal}
  {\bibinfo  {journal} {Phys. Rev. B}\ }\textbf {\bibinfo {volume} {50}},\
  \bibinfo {pages} {1760} (\bibinfo {year} {1994})}\BibitemShut {NoStop}%
\bibitem [{\citenamefont {Yang}\ \emph {et~al.}(2021)\citenamefont {Yang},
  \citenamefont {Zibrov}, \citenamefont {Bai}, \citenamefont {Taniguchi},
  \citenamefont {Watanabe}, \citenamefont {Zaletel},\ and\ \citenamefont
  {Young}}]{yang_experimental_2021}%
  \BibitemOpen
  \bibfield  {author} {\bibinfo {author} {\bibfnamefont {F.}~\bibnamefont
  {Yang}}, \bibinfo {author} {\bibfnamefont {A.~A.}\ \bibnamefont {Zibrov}},
  \bibinfo {author} {\bibfnamefont {R.}~\bibnamefont {Bai}}, \bibinfo {author}
  {\bibfnamefont {T.}~\bibnamefont {Taniguchi}}, \bibinfo {author}
  {\bibfnamefont {K.}~\bibnamefont {Watanabe}}, \bibinfo {author}
  {\bibfnamefont {M.~P.}\ \bibnamefont {Zaletel}}, \ and\ \bibinfo {author}
  {\bibfnamefont {A.~F.}\ \bibnamefont {Young}},\ }\href {\doibase
  10.1103/PhysRevLett.126.156802} {\bibfield  {journal} {\bibinfo  {journal}
  {Physical Review Letters}\ }\textbf {\bibinfo {volume} {126}},\ \bibinfo
  {pages} {156802} (\bibinfo {year} {2021})},\ \bibinfo {note} {publisher:
  American Physical Society}\BibitemShut {NoStop}%
\bibitem [{\citenamefont {Gammel}\ \emph {et~al.}(1988)\citenamefont {Gammel},
  \citenamefont {Bishop}, \citenamefont {Eisenstein}, \citenamefont {English},
  \citenamefont {Gossard}, \citenamefont {Ruel},\ and\ \citenamefont
  {Stormer}}]{gammel_ultralow-temperature_1988}%
  \BibitemOpen
  \bibfield  {author} {\bibinfo {author} {\bibfnamefont {P.~L.}\ \bibnamefont
  {Gammel}}, \bibinfo {author} {\bibfnamefont {D.~J.}\ \bibnamefont {Bishop}},
  \bibinfo {author} {\bibfnamefont {J.~P.}\ \bibnamefont {Eisenstein}},
  \bibinfo {author} {\bibfnamefont {J.~H.}\ \bibnamefont {English}}, \bibinfo
  {author} {\bibfnamefont {A.~C.}\ \bibnamefont {Gossard}}, \bibinfo {author}
  {\bibfnamefont {R.}~\bibnamefont {Ruel}}, \ and\ \bibinfo {author}
  {\bibfnamefont {H.~L.}\ \bibnamefont {Stormer}},\ }\href {\doibase
  10.1103/PhysRevB.38.10128} {\bibfield  {journal} {\bibinfo  {journal}
  {Physical Review B}\ }\textbf {\bibinfo {volume} {38}},\ \bibinfo {pages}
  {10128} (\bibinfo {year} {1988})},\ \bibinfo {note} {publisher: American
  Physical Society}\BibitemShut {NoStop}%
\bibitem [{\citenamefont {Falson}\ \emph {et~al.}(2015)\citenamefont {Falson},
  \citenamefont {Maryenko}, \citenamefont {Friess}, \citenamefont {Zhang},
  \citenamefont {Kozuka}, \citenamefont {Tsukazaki}, \citenamefont {Smet},\
  and\ \citenamefont {Kawasaki}}]{falson_even-denominator_2015}%
  \BibitemOpen
  \bibfield  {author} {\bibinfo {author} {\bibfnamefont {J.}~\bibnamefont
  {Falson}}, \bibinfo {author} {\bibfnamefont {D.}~\bibnamefont {Maryenko}},
  \bibinfo {author} {\bibfnamefont {B.}~\bibnamefont {Friess}}, \bibinfo
  {author} {\bibfnamefont {D.}~\bibnamefont {Zhang}}, \bibinfo {author}
  {\bibfnamefont {Y.}~\bibnamefont {Kozuka}}, \bibinfo {author} {\bibfnamefont
  {A.}~\bibnamefont {Tsukazaki}}, \bibinfo {author} {\bibfnamefont {J.~H.}\
  \bibnamefont {Smet}}, \ and\ \bibinfo {author} {\bibfnamefont
  {M.}~\bibnamefont {Kawasaki}},\ }\href {\doibase 10.1038/nphys3259}
  {\bibfield  {journal} {\bibinfo  {journal} {Nature Physics}\ }\textbf
  {\bibinfo {volume} {11}},\ \bibinfo {pages} {347} (\bibinfo {year}
  {2015})}\BibitemShut {NoStop}%
\bibitem [{\citenamefont {Shi}\ \emph {et~al.}(2020)\citenamefont {Shi},
  \citenamefont {Shih}, \citenamefont {Gustafsson}, \citenamefont {Rhodes},
  \citenamefont {Kim}, \citenamefont {Watanabe}, \citenamefont {Taniguchi},
  \citenamefont {Papić}, \citenamefont {Hone},\ and\ \citenamefont
  {Dean}}]{shi_odd-_2020}%
  \BibitemOpen
  \bibfield  {author} {\bibinfo {author} {\bibfnamefont {Q.}~\bibnamefont
  {Shi}}, \bibinfo {author} {\bibfnamefont {E.-M.}\ \bibnamefont {Shih}},
  \bibinfo {author} {\bibfnamefont {M.~V.}\ \bibnamefont {Gustafsson}},
  \bibinfo {author} {\bibfnamefont {D.~A.}\ \bibnamefont {Rhodes}}, \bibinfo
  {author} {\bibfnamefont {B.}~\bibnamefont {Kim}}, \bibinfo {author}
  {\bibfnamefont {K.}~\bibnamefont {Watanabe}}, \bibinfo {author}
  {\bibfnamefont {T.}~\bibnamefont {Taniguchi}}, \bibinfo {author}
  {\bibfnamefont {Z.}~\bibnamefont {Papić}}, \bibinfo {author} {\bibfnamefont
  {J.}~\bibnamefont {Hone}}, \ and\ \bibinfo {author} {\bibfnamefont {C.~R.}\
  \bibnamefont {Dean}},\ }\href {\doibase 10.1038/s41565-020-0685-6} {\bibfield
   {journal} {\bibinfo  {journal} {Nature Nanotechnology}\ }\textbf {\bibinfo
  {volume} {15}},\ \bibinfo {pages} {569} (\bibinfo {year} {2020})},\ \bibinfo
  {note} {number: 7 Publisher: Nature Publishing Group}\BibitemShut {NoStop}%
\bibitem [{\citenamefont {Jain}(1989)}]{jain_composite-fermion_1989}%
  \BibitemOpen
  \bibfield  {author} {\bibinfo {author} {\bibfnamefont {J.~K.}\ \bibnamefont
  {Jain}},\ }\href {\doibase 10.1103/PhysRevLett.63.199} {\bibfield  {journal}
  {\bibinfo  {journal} {Physical Review Letters}\ }\textbf {\bibinfo {volume}
  {63}},\ \bibinfo {pages} {199} (\bibinfo {year} {1989})},\ \bibinfo {note}
  {publisher: American Physical Society}\BibitemShut {NoStop}%
\bibitem [{\citenamefont {Levin}\ and\ \citenamefont
  {Halperin}(2009)}]{levin_collective_2009}%
  \BibitemOpen
  \bibfield  {author} {\bibinfo {author} {\bibfnamefont {M.}~\bibnamefont
  {Levin}}\ and\ \bibinfo {author} {\bibfnamefont {B.~I.}\ \bibnamefont
  {Halperin}},\ }\href {\doibase 10.1103/PhysRevB.79.205301} {\bibfield
  {journal} {\bibinfo  {journal} {Physical Review B}\ }\textbf {\bibinfo
  {volume} {79}},\ \bibinfo {pages} {205301} (\bibinfo {year}
  {2009})}\BibitemShut {NoStop}%
\bibitem [{\citenamefont {Eisenstein}\ \emph {et~al.}(1992)\citenamefont
  {Eisenstein}, \citenamefont {Pfeiffer},\ and\ \citenamefont
  {West}}]{eisenstein_negative_1992}%
  \BibitemOpen
  \bibfield  {author} {\bibinfo {author} {\bibfnamefont {J.~P.}\ \bibnamefont
  {Eisenstein}}, \bibinfo {author} {\bibfnamefont {L.~N.}\ \bibnamefont
  {Pfeiffer}}, \ and\ \bibinfo {author} {\bibfnamefont {K.~W.}\ \bibnamefont
  {West}},\ }\href {\doibase 10.1103/PhysRevLett.68.674} {\bibfield  {journal}
  {\bibinfo  {journal} {Phys. Rev. Lett.}\ }\textbf {\bibinfo {volume} {68}},\
  \bibinfo {pages} {674} (\bibinfo {year} {1992})}\BibitemShut {NoStop}%
\bibitem [{\citenamefont {Zaletel}\ \emph {et~al.}(2013)\citenamefont
  {Zaletel}, \citenamefont {Mong},\ and\ \citenamefont
  {Pollmann}}]{zaletel_topological_2013}%
  \BibitemOpen
  \bibfield  {author} {\bibinfo {author} {\bibfnamefont {M.~P.}\ \bibnamefont
  {Zaletel}}, \bibinfo {author} {\bibfnamefont {R.~S.~K.}\ \bibnamefont
  {Mong}}, \ and\ \bibinfo {author} {\bibfnamefont {F.}~\bibnamefont
  {Pollmann}},\ }\href {\doibase 10.1103/PhysRevLett.110.236801} {\bibfield
  {journal} {\bibinfo  {journal} {Physical Review Letters}\ }\textbf {\bibinfo
  {volume} {110}},\ \bibinfo {pages} {236801} (\bibinfo {year} {2013})},\
  \bibinfo {note} {publisher: American Physical Society}\BibitemShut {NoStop}%
\bibitem [{\citenamefont {Mong}\ \emph {et~al.}(2017)\citenamefont {Mong},
  \citenamefont {Zaletel}, \citenamefont {Pollmann},\ and\ \citenamefont
  {Papić}}]{mong_fibonacci_2017}%
  \BibitemOpen
  \bibfield  {author} {\bibinfo {author} {\bibfnamefont {R.~S.~K.}\
  \bibnamefont {Mong}}, \bibinfo {author} {\bibfnamefont {M.~P.}\ \bibnamefont
  {Zaletel}}, \bibinfo {author} {\bibfnamefont {F.}~\bibnamefont {Pollmann}}, \
  and\ \bibinfo {author} {\bibfnamefont {Z.}~\bibnamefont {Papić}},\ }\href
  {\doibase 10.1103/PhysRevB.95.115136} {\bibfield  {journal} {\bibinfo
  {journal} {Physical Review B}\ }\textbf {\bibinfo {volume} {95}},\ \bibinfo
  {pages} {115136} (\bibinfo {year} {2017})},\ \bibinfo {note} {publisher:
  American Physical Society}\BibitemShut {NoStop}%
\bibitem [{\citenamefont {Polyakov}\ and\ \citenamefont
  {Shklovskii}(1995)}]{polyakov_universal_1995}%
  \BibitemOpen
  \bibfield  {author} {\bibinfo {author} {\bibfnamefont {D.~G.}\ \bibnamefont
  {Polyakov}}\ and\ \bibinfo {author} {\bibfnamefont {B.~I.}\ \bibnamefont
  {Shklovskii}},\ }\href {\doibase 10.1103/PhysRevLett.74.150} {\bibfield
  {journal} {\bibinfo  {journal} {Physical Review Letters}\ }\textbf {\bibinfo
  {volume} {74}},\ \bibinfo {pages} {150} (\bibinfo {year} {1995})},\ \bibinfo
  {note} {publisher: American Physical Society}\BibitemShut {NoStop}%
\bibitem [{\citenamefont {d’Ambrumenil}\ \emph {et~al.}(2011)\citenamefont
  {d’Ambrumenil}, \citenamefont {Halperin},\ and\ \citenamefont
  {Morf}}]{dambrumenil_model_2011}%
  \BibitemOpen
  \bibfield  {author} {\bibinfo {author} {\bibfnamefont {N.}~\bibnamefont
  {d’Ambrumenil}}, \bibinfo {author} {\bibfnamefont {B.~I.}\ \bibnamefont
  {Halperin}}, \ and\ \bibinfo {author} {\bibfnamefont {R.~H.}\ \bibnamefont
  {Morf}},\ }\href {\doibase 10.1103/PhysRevLett.106.126804} {\bibfield
  {journal} {\bibinfo  {journal} {Physical Review Letters}\ }\textbf {\bibinfo
  {volume} {106}},\ \bibinfo {pages} {126804} (\bibinfo {year} {2011})},\
  \bibinfo {note} {publisher: American Physical Society}\BibitemShut {NoStop}%
\bibitem [{\citenamefont {Nuebler}\ \emph {et~al.}(2010)\citenamefont
  {Nuebler}, \citenamefont {Umansky}, \citenamefont {Morf}, \citenamefont
  {Heiblum}, \citenamefont {von Klitzing},\ and\ \citenamefont
  {Smet}}]{nuebler_density_2010}%
  \BibitemOpen
  \bibfield  {author} {\bibinfo {author} {\bibfnamefont {J.}~\bibnamefont
  {Nuebler}}, \bibinfo {author} {\bibfnamefont {V.}~\bibnamefont {Umansky}},
  \bibinfo {author} {\bibfnamefont {R.}~\bibnamefont {Morf}}, \bibinfo {author}
  {\bibfnamefont {M.}~\bibnamefont {Heiblum}}, \bibinfo {author} {\bibfnamefont
  {K.}~\bibnamefont {von Klitzing}}, \ and\ \bibinfo {author} {\bibfnamefont
  {J.}~\bibnamefont {Smet}},\ }\href
  {http://link.aps.org/doi/10.1103/PhysRevB.81.035316} {\bibfield  {journal}
  {\bibinfo  {journal} {Phys. Rev. B}\ }\textbf {\bibinfo {volume} {81}}
  (\bibinfo {year} {2010})}\BibitemShut {NoStop}%
\bibitem [{\citenamefont {Ezawa}\ and\ \citenamefont
  {Iwazaki}(1992)}]{ezawa_fractional_1992}%
  \BibitemOpen
  \bibfield  {author} {\bibinfo {author} {\bibfnamefont {Z.~F.}\ \bibnamefont
  {Ezawa}}\ and\ \bibinfo {author} {\bibfnamefont {A.}~\bibnamefont
  {Iwazaki}},\ }\href {\doibase 10.1143/JPSJ.61.4133} {\bibfield  {journal}
  {\bibinfo  {journal} {Journal of the Physical Society of Japan}\ }\textbf
  {\bibinfo {volume} {61}},\ \bibinfo {pages} {4133} (\bibinfo {year}
  {1992})},\ \bibinfo {note} {publisher: The Physical Society of
  Japan}\BibitemShut {NoStop}%
\bibitem [{\citenamefont {Ma}\ \emph {et~al.}(2022)\citenamefont {Ma},
  \citenamefont {Peterson}, \citenamefont {Scarola},\ and\ \citenamefont
  {Yang}}]{ma_fractional_2022}%
  \BibitemOpen
  \bibfield  {author} {\bibinfo {author} {\bibfnamefont {K.~K.~W.}\
  \bibnamefont {Ma}}, \bibinfo {author} {\bibfnamefont {M.~R.}\ \bibnamefont
  {Peterson}}, \bibinfo {author} {\bibfnamefont {V.~W.}\ \bibnamefont
  {Scarola}}, \ and\ \bibinfo {author} {\bibfnamefont {K.}~\bibnamefont
  {Yang}},\ }\href {\doibase 10.48550/arXiv.2208.07908} {\enquote {\bibinfo
  {title} {Fractional quantum {Hall} effect at the filling factor
  \${\textbackslash}nu=5/2\$},}\ } (\bibinfo {year} {2022}),\ \bibinfo {note}
  {arXiv:2208.07908 [cond-mat]}\BibitemShut {NoStop}%
\bibitem [{\citenamefont {Shizuya}(2007)}]{shizuya_electromagnetic_2007}%
  \BibitemOpen
  \bibfield  {author} {\bibinfo {author} {\bibfnamefont {K.}~\bibnamefont
  {Shizuya}},\ }\href {http://link.aps.org/doi/10.1103/PhysRevB.75.245417}
  {\bibfield  {journal} {\bibinfo  {journal} {Phys. Rev. B}\ }\textbf {\bibinfo
  {volume} {75}} (\bibinfo {year} {2007})}\BibitemShut {NoStop}%
\bibitem [{\citenamefont {Cooper}\ and\ \citenamefont
  {Stern}(2009)}]{cooper_observable_2009}%
  \BibitemOpen
  \bibfield  {author} {\bibinfo {author} {\bibfnamefont {N.~R.}\ \bibnamefont
  {Cooper}}\ and\ \bibinfo {author} {\bibfnamefont {A.}~\bibnamefont {Stern}},\
  }\href {\doibase 10.1103/PhysRevLett.102.176807} {\bibfield  {journal}
  {\bibinfo  {journal} {Physical Review Letters}\ }\textbf {\bibinfo {volume}
  {102}},\ \bibinfo {pages} {176807} (\bibinfo {year} {2009})}\BibitemShut
  {NoStop}%
\bibitem [{Note1()}]{Note1}%
  \BibitemOpen
  \bibinfo {note} {A. Yazdani, Private communication}\BibitemShut {NoStop}%
\bibitem [{\citenamefont {Geick}\ \emph {et~al.}(1966)\citenamefont {Geick},
  \citenamefont {Perry},\ and\ \citenamefont {Rupprecht}}]{geick_normal_1966}%
  \BibitemOpen
  \bibfield  {author} {\bibinfo {author} {\bibfnamefont {R.}~\bibnamefont
  {Geick}}, \bibinfo {author} {\bibfnamefont {C.~H.}\ \bibnamefont {Perry}}, \
  and\ \bibinfo {author} {\bibfnamefont {G.}~\bibnamefont {Rupprecht}},\ }\href
  {\doibase 10.1103/PhysRev.146.543} {\bibfield  {journal} {\bibinfo  {journal}
  {Physical Review}\ }\textbf {\bibinfo {volume} {146}},\ \bibinfo {pages}
  {543} (\bibinfo {year} {1966})}\BibitemShut {NoStop}%
\bibitem [{\citenamefont {Misumi}\ and\ \citenamefont
  {Shizuya}(2008)}]{misumi_electromagnetic_2008}%
  \BibitemOpen
  \bibfield  {author} {\bibinfo {author} {\bibfnamefont {T.}~\bibnamefont
  {Misumi}}\ and\ \bibinfo {author} {\bibfnamefont {K.}~\bibnamefont
  {Shizuya}},\ }\href {\doibase 10.1103/PhysRevB.77.195423} {\bibfield
  {journal} {\bibinfo  {journal} {Physical Review B}\ }\textbf {\bibinfo
  {volume} {77}},\ \bibinfo {pages} {195423} (\bibinfo {year}
  {2008})}\BibitemShut {NoStop}%
\bibitem [{\citenamefont {Maki}\ and\ \citenamefont
  {Zotos}(1983)}]{maki_static_1983}%
  \BibitemOpen
  \bibfield  {author} {\bibinfo {author} {\bibfnamefont {K.}~\bibnamefont
  {Maki}}\ and\ \bibinfo {author} {\bibfnamefont {X.}~\bibnamefont {Zotos}},\
  }\href {\doibase 10.1103/PhysRevB.28.4349} {\bibfield  {journal} {\bibinfo
  {journal} {Physical Review B}\ }\textbf {\bibinfo {volume} {28}},\ \bibinfo
  {pages} {4349} (\bibinfo {year} {1983})},\ \bibinfo {note} {publisher:
  American Physical Society}\BibitemShut {NoStop}%
\end{thebibliography}
\end{document}